\documentclass[ba]{imsart}

\RequirePackage{amsthm,amsmath,amsfonts,amssymb}
\RequirePackage[numbers]{natbib}
\RequirePackage[colorlinks,citecolor=blue,urlcolor=blue,backref=page,backref=page]{hyperref}
\RequirePackage{graphicx}
\usepackage{algorithm}
\usepackage{algpseudocode}
\usepackage{xcolor}
\usepackage{multirow}
\usepackage{adjustbox}
\usepackage{siunitx}
\usepackage{framed}
\usepackage{booktabs}
\usepackage[normalem]{ulem}
\firstpage{1}
\lastpage{1}
  
\startlocaldefs
\theoremstyle{plain}

\theoremstyle{definition}

\theoremstyle{remark}

\endlocaldefs

\begin{document}

\begin{frontmatter}
\title{Approximating evidence via bounded harmonic means\thanksref{rip}}
\runtitle{Harmonic means for evidence}
\thankstext{rip}{This paper is dedicated to the memory of Karim Benabed, who collaborated with CPR and DW on evidence approximation in \cite{kilbinger2010evidence} and who died in a cycling accident in December 2025.}

\begin{aug}

\author[A]{\fnms{Dana}~\snm{Naderi}\ead[label=e1]{naderi@ceremade.dauphine.fr}},
\author[A,B]{\fnms{Christian P}~\snm{Robert}\ead[label=e2]{xian@ceremade.dauphine.fr}\orcid{0000-0001-6635-3261}},
\author[C]{\fnms{Kaniav}~\snm{Kamary}\ead[label=e3]{kaniav.kamary@insa-lyon.fr}},\and
\author[D]{\fnms{Darren}~\snm{Wraith}\ead[label=e4]{d.wraith@qut.edu.au}\orcid{0000-0001-8755-6471}}
\address[A]{CEREMADE, Université Paris Dauphine-PSL \printead[presep={,\ }]{e1}}
\address[B]{Department of Statistics, University of Warwick\printead[presep={,\ }]{e2}}
\address[C]{Institut Camille Jordan,
INSA Lyon\printead[presep={,\ }]{e3}}
\address[D]{School of Public Health \& Social Work and Centre for Data Science,
Queensland University of Technology\printead[presep={,\ }]{e4}}
\end{aug}

\begin{abstract}
Efficient Bayesian model selection relies on the model evidence or marginal likelihood, whose computation often requires evaluating an intractable integral. The harmonic mean estimator (HME) has long been a standard method of approximating the evidence.  
While computationally simple, the version introduced by \citet{newton1994approximate} potentially suffers from infinite variance. To overcome this issue, \citet{gelfand1994bayesian} defined a standardized representation of the estimator based on an instrumental function and \citet{robert2009computational} later proposed to use higher posterior density (HPD) indicators as instrumental functions. 
 Following this approach, a practical method is proposed, based on an elliptical covering of the HPD region with non-overlapping ellipsoids. The resulting estimator, called the Elliptical Covering Marginal Likelihood Estimator (ECMLE), not only eliminates the infinite-variance issue of the original HME and allows exact volume computations, 
 but is also able to be used in multimodal settings.
Through several examples, we illustrate that ECMLE outperforms other recent methods such as THAMES and its improved version \citep{metodiev2025easily}. 
Moreover, ECMLE demonstrates lower variance—a key challenge that subsequent HME variants have sought to address—and provides more stable evidence approximations, even in challenging settings.
\end{abstract}


\end{frontmatter}

\section{Introduction}\label{sec1}
In Bayesian inference, the marginal likelihood allows for the assessment of how well a model explains the observed data, setting the foundation for Bayesian model comparison.
Bayesian model selection 
proceeds by computing Bayes factors for pairs of models \citep{jeffreys1998theory,robert2009harold}, which quantifies the evidence in favor of one model over the other as the ratio of their respective marginal likelihoods. The marginal likelihood measures the average fit of a model to the observed data by evaluating the expectation of the likelihood under the prior. In practice, the expectation integral is rarely available in closed form, inducing a major computational bottleneck in Bayesian inference.

To tackle this computational issue, various Monte Carlo–based estimators have been developed \citep{marin2010montecarlo}, including Chib’s estimator \citep{chib1995marginal}, Bridge Sampling \citep{meng1996simulating}, importance sampling (IS) \citep{geweke1989bayesian}, and other related techniques, offering diverse approximations to the marginal likelihood.
These approaches, however, often require significant computational resources or impose restrictive assumptions.
 In particular, the IS may perform poorly when the posterior is multimodal, due to difficulty in adequately covering all modes with a single proposal distribution \citep{diciccio1997computing}.

In this collection, the Harmonic Mean Estimator (HME) stands out. Proposed by \citet{newton1994approximate}, 
the estimator approximates the evidence based on a sample from the posterior distribution $\pi(\theta|x)$ and only requires the computation of the likelihood $L(\theta|x)$ for its implementation. It is, nevertheless, often unstable and tends to fail, especially when dealing with complex models characterized by heavy-tailed likelihoods or irregular posterior structures, as shown by \citet{neal1994contribution} who pointed out the dangers of relying solely on HME. In parallel, \citet{gelfand1994bayesian} proposed a general identity 
\begin{equation}
\int \dfrac{\varphi(\theta)}{\pi(\theta)L(\theta|x)}\pi(\theta|x)\,\text d\theta = 1\Big/ \int \pi(\theta)L(\theta|x) \,\text d\theta
\label{gelfandey}
\end{equation}
for expanding HMEs (known as the Gelfand–Dey estimator) by allowing for more flexible instrumental density functions $\varphi(\theta)$ than the prior density $\pi(\theta)$. The Gelfand-Dey's methodology further applies to complex models  including hierarchical and non-conjugate models. However, the accuracy of their estimators heavily depends on the choice of the instrumental function $\varphi(\theta)$. As a result, \citet{robert2009computational} and \citet{marin2010montecarlo} later introduced HPD-truncated estimators based on the representation \eqref{gelfandey}, which select the instrumental density as a uniform distribution over approximate highest posterior density (HPD) regions. 
The authors based their approach on the convex hull of MCMC simulations with the highest density values, ensuring boundedness of the importance ratio in \eqref{gelfandey} but requiring a computationally costly derivation of the convex hull.
Similarly, \citet{10.1214/12-BA725} 
used an indicator function as part of the instrumental function so as to limit the integration domain to a well-sampled region, thereby improving the stability and accuracy of the marginal likelihood estimators. However, \citet{10.1214/12-BA725}'s approach 
becomes computationally intensive for complex posteriors such as heavy tail or multimodal cases.
\citet{wang2017new} proposed the Partition Weighted Kernel (PWK) estimator, based on  a weighted sum of indicator functions over a partition of the parameter space leading to more stable and consistent estimates of the evidence. The ensuing method is however more computationally intensive than a simple estimator since it requires a density estimation within each partition and adds costly post-processing. Furthermore, in the case of complex posteriors, the partitioning of the space may fail to capture accurately the structure of the posterior.

As an alternative, \citet{caldwell2020integration} introduced a practical partitioning scheme based on multiple non-overlapping hypercubes, which proves effective in reducing the instability of harmonic mean estimators in many practical settings. 
However, as the parameter dimension increases, it becomes exponentially harder to find well-sampled and bounded hypercubes that are significant for the target distribution.
In another alternative approach, \citet{mcewen2021machine} introduced the learnt harmonic mean estimator, building upon \citet{gelfand1994bayesian}'s reciprocal importance sampling framework that uses machine learning models to approximate the optimal target distribution. \citet{mcewen2021machine}'s approach is, however, computationally expensive, requiring costly density estimation, and depending on the quality of the learned target distribution, where poor approximations can lead to instability or suboptimal performance.

Closer to our proposal, \citet{reichl2020estimating} constructed a geometrically motivated marginal likelihood estimator that leverages posterior draws and an ellipsoidal region centered on the posterior mode (MAP) 
to compute the evidence stably and efficiently, i.e., avoiding the high variance of classical estimators.
The ellipsoid central to the method is derived from the empirical covariance of the posterior draws and most crucially, it enables an exact analytic calculation of its volume.
More recently, \citet{metodiev2024easily} adopted a similar approach they called the Truncated Harmonic Mean Estimator (THAMES), where the ellipsoid is derived from a Gaussian distribution centered at the posterior mean or MAP, while within
approximate $\alpha$-HPD regions. The covariance matrix behind the ellipsoid is estimated from an MCMC posterior sample and the method seeks to minimize the estimator’s variance by selecting an optimal ellipsoid radius to balance bias versus variance. 
Due to the simple and closed form of their estimators, \citet{reichl2020estimating}'s and \citet{metodiev2024easily}'s methods are practical and easy to implement. Their application is, however, limited to unimodal, smooth, and roughly Gaussian-shaped posteriors with low to moderate dimensions. Indeed, in multimodal or strongly skewed posterior cases, the ellipsoidal truncation may miss significant posterior regions and  include irrelevant, low-posterior-density areas. To overcome these difficulties, \citet{metodiev2025easily} subsequently adapted THAMES for multivariate mixture models, addressing key limitations of the earlier method, but the efficiency of their approach still depends on adaptation to the geometry of the posterior region and a potential computational cost arising from using auxiliary simulations to evaluate complex intersection volumes.

In the current paper, we expand on the works of \citet{wraith2009estimation}, \citet{robert2009computational}, and \citet{metodiev2024easily} to develop a flexible approach based on multiple elliptical coverings for marginal likelihood estimation (under the acronym ECMLE). ECMLE aims to approximate an HPD region of the posterior distribution via a simulation-based union of non-overlapping ellipsoids. 
We demonstrate the practical advantages of the estimator through several examples, including multivariate Gaussians, mixtures of multivariate Gaussians, and a Rosenbrock distribution. The results show that the ECMLE method significantly reduces the variance of the marginal likelihood estimates, providing reliable approximations even in challenging scenarios. This advancement in evidence approximation opens up new possibilities for accurately evaluating Bayesian models, particularly in complex posterior landscapes.

The plan of the paper is as follows. Section \ref{sec:overview} introduces the framework for approximating the evidence and reviews alternative harmonic mean approaches proposed in the literature.
Section \ref{sec:ECMLE} precisely describes the ECMLE algorithm along with its implementation details and theoretical justifications. 
Section \ref{sec:illust} illustrates the method performance through several simulation experiments in increasingly complex models. Section \ref{sec:discus} concludes the paper with some discussion on issues and further research.

\section{Approximating the evidence by harmonic means}\label{sec:overview}
Suppose that $\pmb{x}$ is a sample of independent and identically distributed random variables, with a probability distribution within a family of distributions parameterized by $\theta$. 
The marginal likelihood is defined as   
\begin{equation*}
    Z = \int \pi(\theta) L(\theta) \, \text d\theta, \label{zk}
\end{equation*}
where $\pi(\theta)$ is the prior density and $L(\theta)$ denotes the likelihood function, omitting $\pmb{x}$ for brevity's sake. The Gelfand–Dey identity \eqref{gelfandey} thus writes as
\begin{equation*}
   \mathbb{E}^{\pi} \left[ \frac{\varphi(\theta)}{\pi(\theta)L(\theta)} \Bigg| \pmb{x} \right] 
   = \int \frac{\varphi(\theta)}{\pi(\theta)L(\theta)} \times \frac{\pi(\theta)L(\theta)}{Z} \, \text d\theta 
   = \dfrac{1}{Z}
\end{equation*}
and it holds for all probability density functions $\varphi(\cdot)$ that are defined over the posterior support. This identity justifies the following estimator
\begin{equation}
   \hat{Z}^{-1} = \frac{1}{T} \sum_{t=1}^{T} \frac{\varphi(\theta^{(t)})}{\pi(\theta^{(t)}) L(\theta^{(t)})}\,,
   \label{zhat}
\end{equation}
where $(\theta^{(1)}, \ldots, \theta^{(T)})$ is a $T$-sample simulated from the posterior distribution $\pi(\cdot|\pmb{x})$, either independently or via MCMC methods \citep{robertcasella2004}. This estimator is unbiased for $Z^{-1}$ and an appropriate choice of $\varphi(\cdot)$ can lead to a finite variance and better numerical properties \citep{robert2009computational}, compared with the original choice $\varphi(\cdot)=\pi(\cdot)$ of \citet{newton1994approximate}.  In the following, we overview some of the density functions $\varphi(\cdot)$ proposed in the literature. 
\subsubsection*{Instrumental prior distribution}
As mentioned earlier, the original implementation of the identity \eqref{gelfandey} corresponds to $\varphi(\cdot)= \pi(\cdot)$, since
\begin{equation*}
   \mathbb{E}^{\pi} \left[ \frac{1}{L(\theta)} \Bigg| \pmb{x} \right] = \int \frac{\pi(\theta)}{Z} \, \text d\theta  = \frac{1}{Z}\,.
   \label{ohme1}
\end{equation*}
The estimator of \citet{newton1994approximate} is thus written as a special case of \eqref{zhat}:
\begin{equation*}
   \hat{Z}^{-1} = \frac{1}{T} \sum_{t=1}^{T} \frac{1}{L(\theta^{(t)})}, 
   \label{ohme2}
\end{equation*}
\citet{neal1994contribution} discussed the shortcomings of this estimator in terms of high or possibly infinite variance. Alternatives are thus clearly needed and, as noted in
\citet{robert2009computational}, $\pi(\cdot)L(\cdot)$ should have fatter tails than  $\varphi(\cdot)$. For instance, this is the case when $\varphi$ has support within a non-trivial HPD region.

\subsubsection*{Gaussian instrumental distribution}\label{MVN1}
\citet{diciccio1997computing} considered choosing $\varphi(\theta) = \mathcal{N}(\theta; \hat{\theta}, \hat{\Sigma})$, a Gaussian distribution with $\hat{\theta}$ and $\hat{\Sigma}$ being the posterior point estimates of the mean and covariance matrix, respectively. 
This choice is rather standard when approximating the posterior but for 
many problems, Gaussian tails may prove insufficiently heavy. To address this issue, \citet{diciccio1997computing} subsequently proposed truncating the proposal distribution. They suggest replacing the Gaussian distribution with $\varphi(\theta) = \mathcal{N}^+_A(\hat{\theta}, \hat{\Sigma})$, a truncated Gaussian distribution restricted to a highest-density ellipsoid with radius $r$ (set as the \textit{truncation} parameter) and defined as
\begin{equation}
   A= \{\theta:(\theta-\hat{\theta})^T \hat{\Sigma}^{-1}(\theta-\hat{\theta})< r^2\}\,.
   \label{trancate}
\end{equation}
The volume can be analytically calculated using
\begin{equation}
   V(A)= \frac{\pi^{d/2} r^d |\hat{\Sigma}|^{\frac{1}{2}}}{\Gamma(d/2 + 1)}, 
   \label{ellipsoidvolume}
\end{equation}
and the normalization constant of the $\mathcal{N}^+_A(\hat{\theta}, \hat{\Sigma})$ density is also available, since $(\theta-\hat{\theta})^T \hat{\Sigma}^{-1}(\theta-\hat{\theta})$ is a chi-squared random variate under $\varphi(\cdot)$.
In the smoothest cases, as for unimodal posteriors, restricting the support of the distribution guarantees that the estimator
$Z^{-1}$ exhibits finite variance and \citet{diciccio1997computing} observed that this truncation enhances the performance of the estimator. However, the method is not necessarily suited for more generic densities and may suffer from the same issues
as the original estimator.

\subsubsection*{Uniform instrumental distribution for a convex hull of $\alpha$-HPD samples}\label{main}
In order to bound the variance of the estimator \eqref{zhat}, \citet{robert2009computational} proposed choosing the instrumental function $\varphi(\cdot)$ as a uniform distribution over a HPD region of coverage $\alpha$ (called an $\alpha-$HPD region). This region is approximated from the $\alpha$ fraction of a simulated posterior sample that corresponds to the largest arguments of $\pi(\theta)L(\theta)$ by constructing a convex hull, $\mathfrak{H}_\alpha$. 
The associated estimator \eqref{zhat} is then
\begin{equation}
   \hat{Z}^{-1} 
   =\frac{1}{T\,V(\mathfrak{H}_\alpha)} \sum_{t=1}^{T} \frac{\mathbb{I}_{\mathfrak{H}_\alpha}(\theta^{(t)})}{\pi(\theta^{(t)}) L(\theta^{(t)})}. 
   \label{hmech}
\end{equation}
The authors illustrated the efficiency of the method on a $2-$dimensional toy example. However, the extension to higher dimensions is uncertain, due to the computational challenge of deriving the volume of the convex hull. In addition, the inclusion of the hull into an HPD region is not guaranteed and may be inefficient. 

\subsubsection*{Partition Weighted Kernel (PWK) instrumental distribution}
\citet{wang2017new} introduced the partition weighted kernel (PWK) estimator, which generalizes the harmonic mean (HM) and inflated density ratio (IDR) methods via local weighting over a partition of the parameter space. Let $\Omega \subset \Theta$ be a working domain where the kernel $q(\theta) = \pi(\theta)L(\theta)$ is bounded away from zero, and let $\{A_1,\dots,A_K\}$ partition $\Omega$ with weights $w_k > 0$. The instrumental function is:
$$\varphi(\theta) = \sum_{k=1}^{K} w_k \mathbf{1}\{\theta \in A_k\}.$$
The PWK estimator is then:
$$\hat{Z}^{-1} = \frac{\frac{1}{T} \sum_{t=1}^{T} \frac{\varphi(\theta^{(t)})}{q(\theta^{(t)})}}{\sum_{k=1}^{K} w_k V(A_k)},$$
where $V(A_k)$ is the volume of $A_k$. Variance is minimized when $q(\theta)$ is nearly constant on each $A_k$, achieved in practice by standardizing the MCMC sample and using concentric spherical shells with representative kernel values at shell midpoints. The estimator is consistent and has finite variance under mild conditions. However, it becomes computationally intensive in high dimensions due to volume calculations and partition refinement, resembling a grid-based approximation.

To handle multimodal or skewed posteriors where constant weights over spherical shells may be inefficient, they proposed an extended version (ePWK) that allows functional weights \( w_k(\theta) \) varying within each partition subset. In practice, this is achieved by further slicing each spherical shell \( A_k \) into \( m_k \) angular subsets \( \{A_{k1}, \dots, A_{k m_k}\} \) and assigning a representative kernel value \( q(\theta^{*}_{k\ell}) \) to each slice. This additional partitioning reduces heterogeneity of the kernel within each subset, improving estimator efficiency for complex posterior geometries, though the computational overhead limits its practicality to lower-dimensional settings.

\subsubsection*{Uniform instrumental distribution over a Gaussian ellipsoid}
Related to the proposal of \citet{diciccio1997computing}, \citet{metodiev2024easily} developed an instrumental distribution called \textit{Truncated Harmonic Mean Estimator} (with the acronym THAMES). THAMES is a combination of both methods previously mentioned and uses a uniform distribution over the ellipsoid $A$ defined in \eqref{trancate}. Since the volume of the ellipsoid $A$ is available as \eqref{ellipsoidvolume}, THAMES bypasses both the computing issue of the volume and the derivation of the convex hull of \citet{robert2009computational}.
Like the previous estimators, THAMES provides an unbiased estimator of $Z^{-1}$, provided that the posterior density does not vanish within the ellipsoid $A$.
While the choice of the radius $r$ towards minimizing the variance of $\hat{Z}^{-1}$ is argued to be about $\sqrt{d+1}$ as in the Gaussian case, there is no guarantee that $A$ is included within a HPD region. Therefore, it faces limitations in multimodal cases (a case illustrated later on Figure \ref{EX2_ellipsoids_created}), which may invalidate THAMES' consistency and reliability in these settings.

\subsubsection*{Uniform instrumental distribution over a truncated Gaussian ellipsoid}
Most recently, \citet{metodiev2025easily} proposed a new version of THAMES, which we denote by tTHAMES (truncated THAMES) for simplicity, and is specifically tailored for multimodal posterior distributions. This updated method relies on \citet{reichl2020estimating}'s proposal and includes a truncation strategy, towards using a support $
A'_{\hat{\theta}, \hat{\Sigma}, r, \alpha}
$
made of the intersection of the ellipsoid $A$ in \eqref{trancate}, and a $50\%$-HPD region. The THAMES for a mixture of $j=1,\ldots, J$ distributions of the same family,  
$$
X_i|\pmb{p},\pmb{\nu} \overset{i.i.d}{\sim} \sum_{j=1}^J p_j f(|\nu_j), \quad i=1,\ldots, n, \quad X_i\in \mathbb{R}^d, \quad \sum_{j=1}^J p_j=1, \quad 0<p_j<1
$$
is defined as
$$
\hat{Z}^{-1} = \frac{1}{J!} \sum_{s \in S} 
\frac{1}{T/2} \sum_{\substack{t = T/2 + 1, \\ \theta^{(t)} \in A'_{\hat{\theta}, \hat{\Sigma}, r, \alpha}}}^{T}
\frac{1/\mathrm{Vol}(A'_{\hat{\theta}, \hat{\Sigma}, r, \alpha})}{L(\theta^{(t)}) \, \pi(\theta^{(t)})}
$$
where $\theta=(\nu_1, \ldots, \nu_J, p_1, \ldots, p_{J-1})$ and $S$ is a selected set of label permutations whose output does not contradict the constraints that hold in 
\begin{equation}
A'_{\hat{\theta}, \hat{\Sigma}, r, \alpha}=\{\theta:(\theta-\hat{\theta})^T \hat{\Sigma}^{-1}(\theta-\hat{\theta})< r^2, \pi(\theta)L(\theta)>\hat{q}_{\alpha}\}
  \label{intersect}  
\end{equation}
and $\hat{q}_{\alpha}$ is the empirical $(1-\alpha)-$quantile of the sample of unnormalized log-posterior values.

While this modification effectively filters out samples with lower likelihoods, ensuring that the importance weights are bounded, the efficiency of this THAMES proposal fundamentally relies on the instrumental set recovering in sufficient 
volume of the HPD of the target posterior. We contend that in some cases (as illustrated in Figure \ref{EX2_ellipsoids_created}), this specific truncation, even with its benefits, may inadvertently exclude significant modes or important regions of the posterior. Furthermore, the method relies on additional Monte Carlo simulations to compute the volume of this complex intersection region, meaning the overall efficiency of the approach is directly tied to the number of these auxiliary simulations, while inducing a bias other methods do not face.

\section{ECMLE method}\label{sec:ECMLE}
We henceforth provide an efficient approach for estimating the marginal likelihood using harmonic mean estimators, relying on a geometric approximation of a posterior high-probability region by ellipsoids. This approach addresses key limitations of earlier solutions by constructing a simple geometric representation of the HPD approximation that accurately captures its structure and is computationally efficient.

\subsubsection*{Overview and motivation}
The ECMLE estimator is based on the Gelfand-Dey identity \eqref{gelfandey} and the framework of \citet{robert2009computational} \eqref{hmech}, where a uniform distribution over a convex hull is used as the instrumental function. In contrast, our method constructs the set $\mathcal{E}$ as a union of disjoint, locally adapted ellipsoids, enabling exact volume computation while effectively capturing multimodal and irregularly shaped posteriors.

The construction proceeds through three steps formalized in Algorithms \ref{alg:step1}--\ref{alg:step3}: sample partitioning for unbiased estimation, adaptive ellipsoid placement, and marginal likelihood computation.

\subsubsection*{Sample partitioning and threshold determination}
Unbiased estimation requires that the region $\mathcal{E}$ is constructed independently of the sample used for evaluation. Given a collection of posterior draws $\{\boldsymbol{\theta}_i\}_{i=1}^{2T}$ from either MCMC or independent sampling, we partition them into two equal subsets. The first subset $\{\boldsymbol{\theta}_t\}_{t=1}^{T}$ is used to construct the disjoint ellipsoid covering of the HPD region, while the second  $\{\boldsymbol{\theta}_t\}_{t=T+1}^{2T}$ evaluates the estimator.

Given the chosen HPD level 
$\alpha$, we determine a threshold 
$c$ on the unnormalized posterior densities $\tilde{\pi}(\theta | x) = \pi(\theta)\, L(x | \theta)$
 that separates the high- and low-density samples (Algorithm~\ref{alg:step1}). This yields the sets of high-density points $\boldsymbol{\Theta}_{\mathrm{HPD}} = \{\boldsymbol{\theta}_t : \tilde{\pi}(\theta_{t}|{x}) \ge c,\ t\le T\}$ and low-density points $\boldsymbol{\Theta}_{\mathrm{LPD}}$. The threshold $c$ serves as the target boundary that ellipsoids must respect.
\begin{algorithm}[h]
\caption{Sample partition and HPD region}
\label{alg:step1}
\begin{algorithmic}[1]
  \State \textbf{Input:} Posterior sample $\{\boldsymbol{\theta}_i\}_{i=1}^{2T}$, unnormalized posterior densities $\{\tilde{\pi}(\theta_{i}|{x})\}_{i=1}^{2T}$, HPD level $\alpha$
  \State \textbf{Output:} Two sample sets and HPD region

  \State Split the sample and its corresponding densities into two parts:
  \[
    \bigl\{\boldsymbol{\theta}_t,\tilde{\pi}(\theta_{t}|{x})\bigr\}_{t=1}^{T}
    \quad\text{and}\quad
    \bigl\{\boldsymbol{\theta}_t,\tilde{\pi}(\theta_{t}|{x})\bigr\}_{t=T+1}^{2T}
  \]

  \State Compute the empirical HPD threshold:
  $
    c \;=\; \mathrm{quantile}\bigl(\{\tilde{\pi}(\theta_{t}|{x})\}_{t=1}^{T},\,1-\alpha\bigr)
  $

  \State Identify sample values within the HPD region for the first part:
  \[
    \boldsymbol{\Theta}_{HPD} = \{\boldsymbol{\theta}_t : \tilde{\pi}(\theta_{t}|{x}) \ge c,\ t\le T\},\quad
    \boldsymbol{\Theta}_{LPD} = \{\boldsymbol{\theta}_t : \tilde{\pi}(\theta_{t}|{x}) < c,\ t\le T\}
  \]

  \State \textbf{Return:} 
    Samples $\{\boldsymbol{\theta}_t\}_{t=1}^{T}$, $\{\boldsymbol{\theta}_t\}_{t=T+1}^{2T}$,
    $\boldsymbol{\Theta}_{HPD}$, and $\boldsymbol{\Theta}_{LPD}$
\end{algorithmic}
\end{algorithm}

\subsubsection*{Adaptive ellipsoid construction}
The core innovation lies in constructing ellipsoids that conform to the local posterior geometry. Rather than imposing a global shape, each ellipsoid adapts to the HPD boundary in its vicinity, maximizing coverage while respecting the threshold $c$. The overall procedure for building this adaptive covering is summarized in Algorithm~\ref{alg:step2}.

\begin{algorithm}[htbp]
\caption{Ellipsoid covering for HPD Regions} 
\label{alg:step2}
\begin{algorithmic}[1]
  \State \textbf{Input:} HPD samples $\boldsymbol{\Theta}_{\text{HPD}}$ with posterior values, LPD samples $\boldsymbol{\Theta}_{LPD}$, posterior function $\pi(\boldsymbol{\theta})$, HPD level $\alpha$
  \State \textbf{Output:} Union of non-overlapping ellipsoids, $\mathcal{E}$, and total volume $V(\mathcal{E})$
  \State Subsample $\boldsymbol{\Theta}_{HPD}$ randomly to size $\lfloor k \cdot |\boldsymbol{\Theta}_{HPD}| \rfloor$
  \State Compute threshold: $c \gets \text{quantile}(\{ \pi(\boldsymbol{\theta}) : \boldsymbol{\theta} \in \boldsymbol{\Theta}_{HPD}\}, 1 - \alpha)$
  \State Order subsampled HPD samples by log-posterior values (highest first)
  \State Initialize: $\mathcal{E} \gets \emptyset$, $V(\mathcal{E}) \gets 0$, $\mathrm{Centers} \gets$ subsampled $\boldsymbol{\Theta}_{HPD}$
  \State $r_{\max} \gets \max_{\boldsymbol{\theta}_i, \boldsymbol{\theta}_j \in \mathrm{Centers}} \|\boldsymbol{\theta}_i - \boldsymbol{\theta}_j\|$
  \While{any $\mathrm{Available}$}
    \State Select $\boldsymbol{\theta}^*$ as next available center (highest log-posterior)
    \State Determine primary direction $\mathbf{u}_1$:
        \State $\boldsymbol{\theta}_{\text{closest}} \gets \arg\min_{\boldsymbol{\theta} \in \boldsymbol{\Theta}_{LPD}} \|\boldsymbol{\theta} - \boldsymbol{\theta}^*\|$
        \State $\mathbf{u}_1 \gets (\boldsymbol{\theta}_{\text{closest}} - \boldsymbol{\theta}^*) \, / \, \|\boldsymbol{\theta}_{\text{closest}} - \boldsymbol{\theta}^*\|$
    \State $r_1 \gets$ bisection solution to $\pi(\boldsymbol{\theta}^* + r \mathbf{u}_1) = c$ over $[0, r_{\max}]$
    \State Compute orthogonal basis $\mathbf{U} \gets$ Gram--Schmidt$([\mathbf{u}_1, \mathbf{e}_2, \dots, \mathbf{e}_d])$
    \State For $i=2$ to $d$:
      \State $\mathbf{u}_i \gets \mathbf{U}_{:,i}$
      \State $r_{+} \gets$ bisection for $+\mathbf{u}_i$; $r_{-} \gets$ bisection for $-\mathbf{u}_i$
      \State $s_i \gets \min(r_{+}, r_{-})$ if valid, else invalid
    \State $s_1 \gets r_1$; $\mathbf{D} \gets \operatorname{diag}(s_1^2, \dots, s_d^2)$; $\boldsymbol{\Sigma} \gets \mathbf{U} \mathbf{D} \mathbf{U}^\top$
    \State $s_{\max} \gets \max(s_1, \dots, s_d)$
    \State Check conservative overlap:
    \If{$\exists \mathfrak{e}_j \in \mathcal{E}$ s.t. $\|\boldsymbol{\theta}^* - \boldsymbol{\mu}_j\| < s_{\max} + s_{\max,j}$}
      \State Mark $\boldsymbol{\theta}^*$ as unavailable; \textbf{continue}
    \EndIf
    \State Compute volume: $v \gets \pi^{d/2} / \Gamma(d/2 + 1) \, \sqrt{\det(\boldsymbol{\Sigma})}$
    \State Add ellipsoid $\mathfrak{e}(\boldsymbol{\theta}^*, \boldsymbol{\Sigma}) = \{\boldsymbol{\theta} : (\boldsymbol{\theta} - \boldsymbol{\theta}^*)^\top \boldsymbol{\Sigma}^{-1} (\boldsymbol{\theta} - \boldsymbol{\theta}^*) \le 1\}$ to $\mathcal{E}$
    \State $V(\mathcal{E}) \gets V(\mathcal{E}) + v$
    \State Prune: for remaining available centers $\boldsymbol{\theta}_k$, if $(\boldsymbol{\theta}_k - \boldsymbol{\theta}^*)^\top \boldsymbol{\Sigma}^{-1} (\boldsymbol{\theta}_k - \boldsymbol{\theta}^*) \le 1$, mark $\boldsymbol{\theta}_k$ as unavailable
    \State Mark $\boldsymbol{\theta}^*$ as unavailable
  \EndWhile
  \State \textbf{Return} $\mathcal{E},\,V(\mathcal{E})$
\end{algorithmic}
\end{algorithm}

We first subsample $\boldsymbol{\Theta}_{\mathrm{HPD}}$ by a rate $k$ (typically 0.05--0.1) to obtain candidate centers, ordered by posterior density to prioritize high-probability regions. For each candidate $\boldsymbol{\theta}^*$, we determine the ellipsoid shape through the following procedure:

A primary axis direction $\mathbf{u}_1$ points toward the nearest low-density point, capturing the dominant direction of posterior decay. Along this axis, we find radius $r_1$ where $\tilde{\pi}(\boldsymbol{\theta}^* + r_1\mathbf{u}_1|x) = c$ via a bisection search. An orthogonal basis $\{\mathbf{u}_1,\ldots,\mathbf{u}_d\}$ is found using Gram-Schmidt, and semi-axes lengths $s_i$ are determined by finding the HPD boundary along each direction, yielding a shape matrix $\boldsymbol{\Sigma} = \mathbf{U}\,\mathrm{diag}(s_1^2,\ldots,s_d^2)\,\mathbf{U}^\top$. The resulting ellipsoid is then defined as
\[
\mathfrak{e}(\boldsymbol{\theta}^*, \boldsymbol{\Sigma}) = \{\boldsymbol{\theta} : (\boldsymbol{\theta} - \boldsymbol{\theta}^*)^\top \boldsymbol{\Sigma}^{-1} (\boldsymbol{\theta} - \boldsymbol{\theta}^*) \le 1\}.
\]

To preserve the non-overlapping property required for volume computation, we adopt a two-step filtering procedure. Candidate centers are processed in order of decreasing posterior density. First, before constructing a new ellipsoid at candidate center $\boldsymbol{\theta}^*$ with maximum semi-axis $s_{\max}$, we check against all existing ellipsoids: if the Euclidean distance to any existing center $\boldsymbol{\mu}_j$ is less than the sum of maximum semi-axes (i.e., $\|\boldsymbol{\theta}^* - \boldsymbol{\mu}_j\| < s_{\max} + s_{\max,j}$), the candidate is rejected. Second, after an ellipsoid $\mathfrak{e}(\boldsymbol{\theta}^*, \boldsymbol{\Sigma})$ is accepted, all remaining candidate centers that fall inside this ellipsoid are pruned from further consideration. This construction guarantees disjointness and enables the exact computation of the total volume as $V(\mathcal{E}) = \sum_j \frac{\pi^{d/2}}{\Gamma(d/2+1)}\sqrt{\det(\boldsymbol{\Sigma}_j)}.$
\subsubsection*{Marginal likelihood computation}
With the ellipsoid collection $\mathcal{E}$ and its total volume $V(\mathcal{E})$ fixed, the marginal likelihood is evaluated using the second sample. For each parameter vector $\{\boldsymbol{\theta}_t\}_{t=T+1}^{2T}$, membership in $\mathcal{E}$ is determined by verifying whether it falls within ellipsoid $\mathfrak{e}_j$. The corresponding estimator is summarized in Algorithm~\ref{alg:step3}.

\begin{algorithm}[h]
\caption{Marginal Likelihood Estimation}
\label{alg:step3}
\begin{algorithmic}[1]
\State \textbf{Input:} Second sample $\{\boldsymbol{\theta}_t\}_{t=T+1}^{2T}$ with unnormalized posterior densities $\{\tilde{\pi}(\theta_{t}|{x})\}_{t=T+1}^{2T}$, ellipsoid collection $\mathcal{E}$, total volume $V(\mathcal{E})$
\State \textbf{Output:} Marginal likelihood estimate $ \hat{Z}$
\State Calculate ECMLE: 
    \[ 
    \hat{Z}^{-1} = \frac{1}{T}\sum_{t=T+1}^{2T}
    \frac{\mathbf{1}_{\mathcal{E}}(\theta_{t})}
     {V(\mathcal{E})\;\tilde\pi(\theta_{t}\mid x)}.
    \] 
\State \textbf{Return:} $\hat{Z}$
\end{algorithmic}
\end{algorithm} 
Note that the dual-purpose roles of the two samples can be exchanged by computing a second estimator of $Z$ that can be averaged with the first one and additionally provides a rough indication of the estimator’s variability. 

\subsubsection*{Computational considerations and complexity}
The computational efficiency of ECMLE is critical for practical applications. We analyze here the time complexity of each algorithm component, demonstrating that the method scales favorably with sample size and dimension.

Let $T$ denote the size of each half-sample (the full posterior sample has size $2T$), 
$d$ the parameter dimension, $\alpha$ the HPD level, 
$k\in(0,1]$ the subsampling rate for HPD candidates, 
$m$ the number of accepted ellipsoids (typically $m\ll T$), 
$T_{\mathrm{HPD}}=\alpha T$, and $T_{\mathrm{LPD}}=(1-\alpha)T$.

For the sample partition and HPD determination step (Algorithm~\ref{alg:step1}), 
computing the threshold and ordering requires sorting the first half-sample:
\[
\text{Time }=O(T\log T).
\]

For the ellipsoid construction stage (Algorithm~\ref{alg:step2}), 
let $kT_{\mathrm{HPD}}$ be the number of candidate centers after subsampling. 
The dominant costs are: 
(i) ordering the candidates, $O(kT_{\mathrm{HPD}}\log(kT_{\mathrm{HPD}}))$; 
and (ii) for each accepted ellipsoid, performing $d$ one-dimensional boundary searches (bisection) 
and overlap checks. 
The resulting time complexity is
\[
\text{Time }=O\Big(kT_{\mathrm{HPD}}\log(kT_{\mathrm{HPD}})\;+\;m\big(T_{\mathrm{LPD}}+d\,J+m\big)\Big),
\]
where $J$ denotes the number of bisection iterations per direction 
(typically small due to the exponential convergence of the method).

Finally, the marginal likelihood evaluation (Algorithm~\ref{alg:step3}) 
requires at most $m$ ellipsoid checks per posterior draw, leading to
\[
\text{Time }=O(m\,T\,d^2).
\]
In typical settings, the subsampling factor $k$ is small (e.g., $0.05$–$0.1$), 
and $m$ remains modest due to pruning, 
so the overall computation scales efficiently even for large $T$.

\section{Numerical Illustrations}\label{sec:illust}

In this section, we empirically compare ECMLE with THAMES estimators through several examples.
We focus on THAMES because it represents the current state of the art among harmonic-mean–based estimators. THAMES and tTHAMES were recently shown to outperform earlier bounded harmonic mean approaches and to provide stable, closed-form evidence approximations across a range of models. As ECMLE builds directly on the same harmonic-mean identity and HPD-based truncation principle, this comparison allows a fair and direct assessment within the same methodological family.

We use the notation $\pmb{x} = \{x_1, \ldots, x_n\}$ to indicate a set of $n$ observations. Each experiment was performed using 100 replications of Algorithms 1--3 for all examples. To ensure a fair comparison across methods, we calibrated the number of posterior draws for each method so that all methods operate under approximately the same computational budget. This approach accounts for the differing per-sample computational costs of each estimator, allowing us to compare estimation accuracy at matched runtime rather than matched sample size.
\subsubsection*{Example 1: Multivariate Gaussian distributions}\label{EX1}
In this first example, we reassess the Multivariate Gaussian case initially considered by \citet{metodiev2024easily}. Take $X_i \in \mathbb{R}^d, i = 1, \ldots, n$, as i.i.d  multivariate Gaussian variables:

\[  
X_i | \pmb{\mu} \overset{}{\sim} \mathcal N_{d}(\pmb{\mu}, I_d), \quad i = 1, \ldots, n,
\] 
where $I_d$ is the $d-$dimensional identity matrix and we choose the following prior distribution for the mean vector $\pmb{\mu}=(\mu_1,\ldots,\mu_d)$:
\[ 
    p(\pmb{\mu}) = \mathcal N_{d}(\pmb{\mu}; 0_d, s I_d),
\] 
with a fixed $s > 0$. The posterior distribution of $\pmb{\mu}$ given the data $\pmb{x}$ is then
\[ 
p(\pmb{\mu} | \pmb{x}) = \mathcal N_{d}(\pmb{\mu}; \hat{\pmb{m}}_{\pmb{\mu} | \pmb{x}}, \hat{s}_{\pmb{\mu} | \pmb{x}} I_d),
\]
where 
\[ 
\hat{\pmb{m}}_{\pmb{\mu} | \pmb{x}} = n\bar{\pmb{x}}/(n + 1/s), \quad \bar{\pmb{x}} = (1/n) \sum_{i=1}^n x_i, \quad \text{and} \quad \hat{s}_{\pmb{\mu} | \pmb{x}} = 1/(n + 1/s).
\] 
As in \citet{metodiev2024easily}, we used a simulated dataset of $n=20$ points with $s=1$, $d=2$ and $\pmb{\mu}=(1,1)$.

\begin{figure}[h!]
\includegraphics[width=0.9\textwidth]{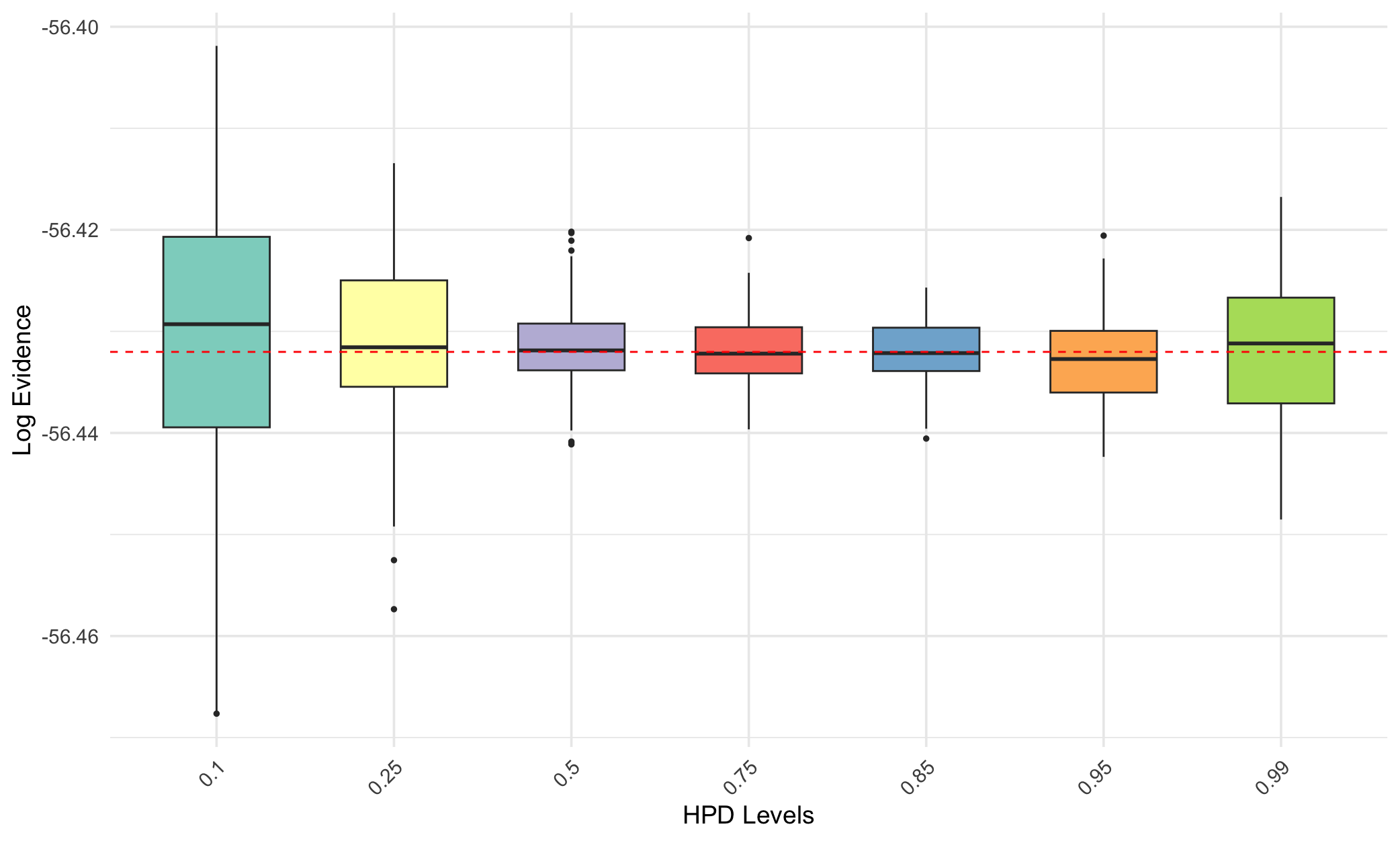}
    \caption{({\bf Example 1}) Comparison of log-marginal-likelihood estimates across different HPD levels using ECMLE. Each box plot was created by performing 100 replications of the algorithms, each using $10^5$ posterior draws for the same observed data. The red dashed line represents the exact value of the marginal likelihood.} 
    \label{Ex1_levels}
\end{figure}

In order to determine the optimal level of our HPD region, we ran $100$ replications of the method for seven different values of HPD levels $\alpha$ from  $10\%$ to $99\%$. The role of the level $\alpha$ is also clear in Figure \ref{Ex1_levels}, as extreme values predictably induce greater variability than when $\alpha$ lies in the upper center of the unit interval. (We stress that the actual coverage of the approximate HPD region $\mathcal E$ is consistently close to its nominal value, for all methods and level choices.)

\begin{figure}[h!]
\includegraphics[width=\textwidth, height=0.3\textheight]{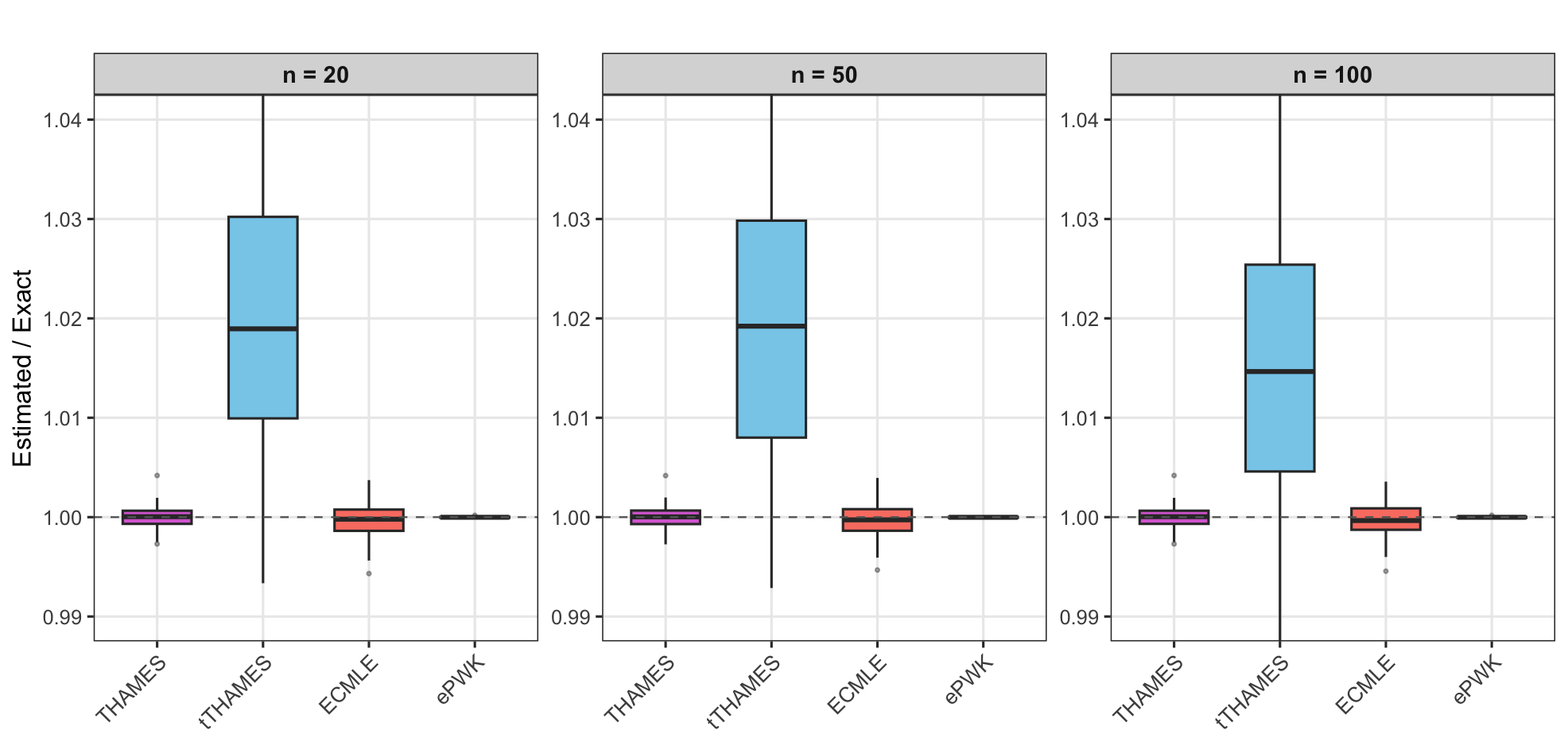}
    \caption{({\bf Example 1}) 
    Evidence ratio $\widehat{Z}/Z$ for the bivariate Gaussian model with $n \in \{20, 50, 100\}$. Each boxplot is based on 100 datasets. To equalize computation time across methods, the number of posterior draws was set to: THAMES (850,000), tTHAMES (38,000), ECMLE (425,000), and ePWK (48,000).
    }
    \label{Ex1_compare}
\end{figure}

In this symmetric and unimodal example, tTHAMES yields poorer results in terms of precision compared to the other estimators, as shown in
Figure \ref{Ex1_compare}. This is likely due to the additional Monte Carlo step required by tTHAMES to estimate the intersection volume between the ellipsoid and the HPD region. 
Furthermore, THAMES is comparable to ECMLE despite the former using an optimal level set and the latter a local model-free approximation. Both THAMES and tTHAMES were implemented using the optimal configurations recommended by \citet{metodiev2024easily,metodiev2025easily}. In THAMES, the ellipsoid radius was set to \( r = \sqrt{d + 1} \), and tTHAMES used the truncation level determined by minimizing the Kolmogorov distance between the truncated, standardized negative log-posterior and the \(\chi^2_d\) distribution. We used the authors' original code for both THAMES and tTHAMES. For tTHAMES, we compute the estimator over the set $A'_{\hat{\theta}, \hat{\Sigma}, r, \alpha}$ (Equation \ref{intersect}), excluding the label-switching corrections for mixture models. Notably, PWK performs well in this setting, benefiting from the low dimensionality and the symmetric, near-circular shape of the posterior, which aligns naturally with its spherical shell partitioning scheme.

\subsubsection*{Example 2: Mixture of multivariate Gaussian distributions}
\label{E2}
We now consider a Gaussian model in $\mathbb R^d$
\[ 
X_i | \pmb{\mu} \overset{\text{iid}}{\sim} \mathcal N(\pmb{\mu}, \Sigma_X), \quad i = 1, \ldots, n,\quad 
\]
with a known value of $\Sigma_X$ and a Gaussian mixture prior on $\pmb{\mu}$ 
\[
p(\pmb{\mu}) = \omega \mathcal{N}(\pmb{\mu} |\pmb{\xi}_1, S_1) + (1 - \omega) \mathcal{N}(\pmb{\mu} | \pmb{\xi}_2, S_2)
\quad 0<\omega<1
\]
The posterior distribution of $\pmb{\mu}$ is then a two-component Gaussian mixture that can be analytically computed as
\[
p(\pmb{\mu} | \pmb{x}) = \hat{\omega} \mathcal{N}(\pmb{\mu} | \pmb{\hat{\xi}}_{n,1}, \hat{S}_{n,1}) + (1-\hat{\omega}) \mathcal{N}(\pmb{\mu} | \pmb{\hat{\xi}}_{n,2}, \hat{S}_{n,2})
\]
where : 
\begin{align*}
\hat{S}_{n,k} &= (n\Sigma_X^{-1} + S_k^{-1})^{-1}\, \quad k=1,2 \\
\pmb{\hat{\xi}}_{n,k} &= \hat{S}_{n,k}(n\Sigma_X^{-1}\bar{\pmb{x}} + S_k^{-1}\pmb{\xi}_k)\\
\hat{\omega} &= \frac{\omega p(\pmb{x}|\pmb{\xi}_1)}{\omega p(\pmb{x}|\pmb{\xi}_1) + (1-\omega) p(\pmb{x}|\pmb{\xi}_2)}
\end{align*}
The exact marginal likelihood is available as
\[
Z = \int p(\pmb{x} | \pmb{\mu}) p(\pmb{\mu})\, \text d\pmb{\mu} = \hat{\omega} ~Z_1 + (1-\hat{\omega}) Z_2
\]
where for $k=1,2$ :
\begin{align*}
Z_k &= (2\pi)^{-\frac{nd}{2}} |\Sigma_X|^{-\frac{(n-1)}{2}} \left|\Sigma_X + n~S_k\right|^{-\frac{1}{2}} \exp\left(-\frac{1}{2}\sum_{i=1}^n (\pmb{x}_i - \bar{\pmb{x}})^T \Sigma_X^{-1} (\pmb{x}_i - \bar{\pmb{x}})\right)\\ &\times \exp\left(-\frac{1}{2}(\bar{x} - \pmb{\xi}_k)^T \left(\frac{1}{n}\Sigma_X + S_k\right)^{-1} (\bar{\pmb{x}} - \pmb{\xi}_k)\right)
\end{align*}

\begin{figure}[h!]
\includegraphics[width=0.3
\textwidth]{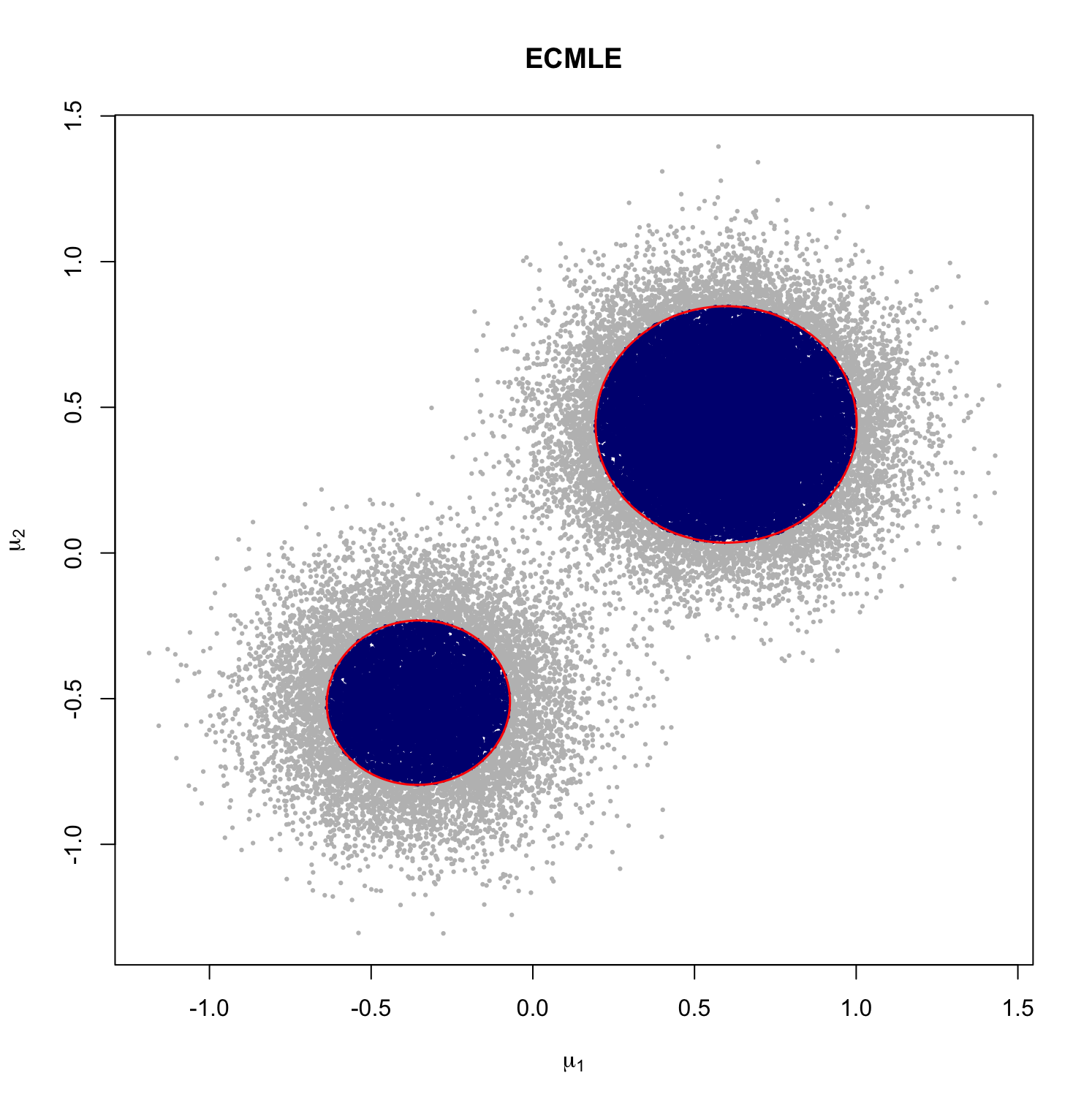}
\includegraphics[width=0.3
\textwidth]{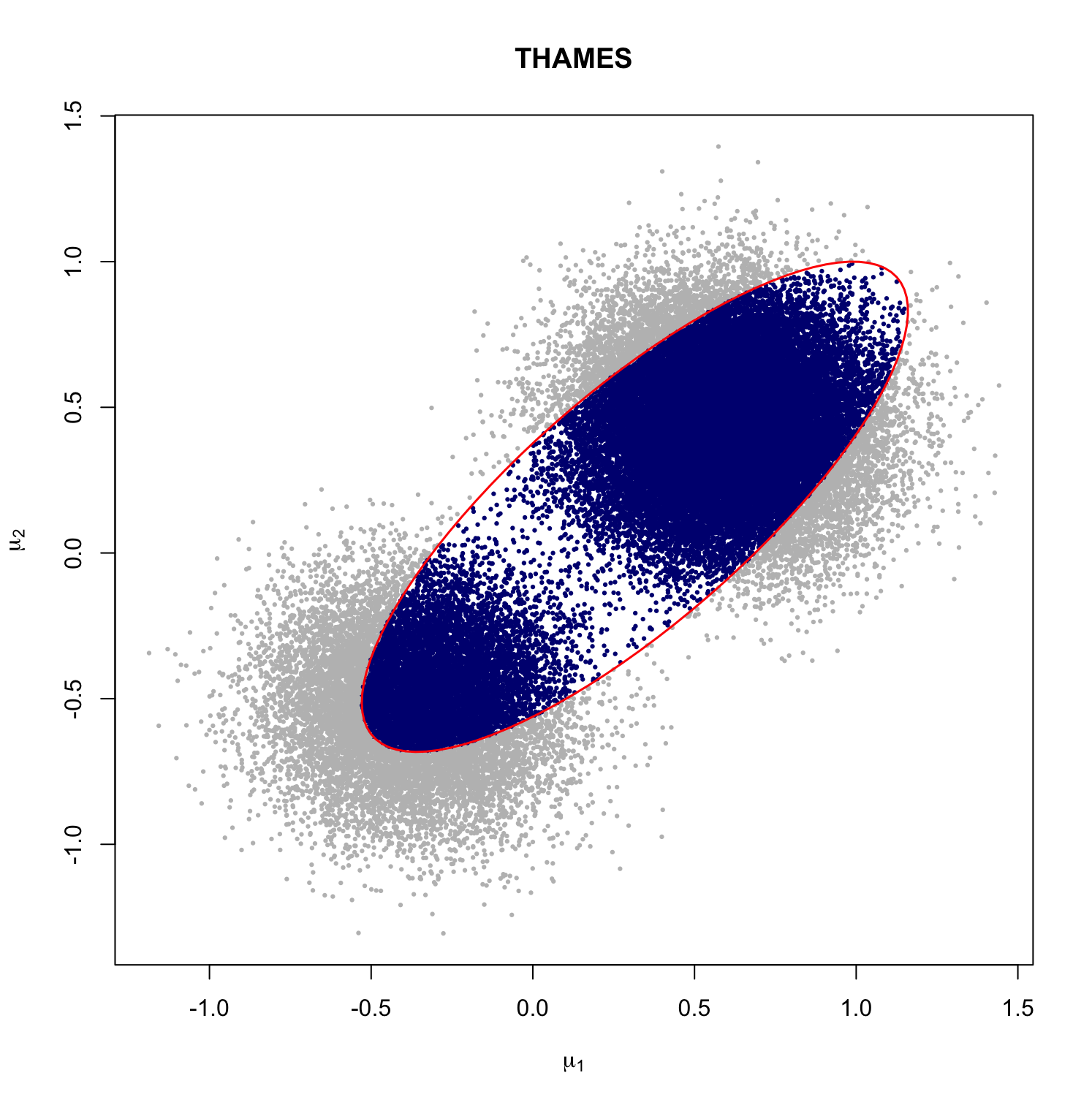}
\includegraphics[width=0.3
\textwidth]{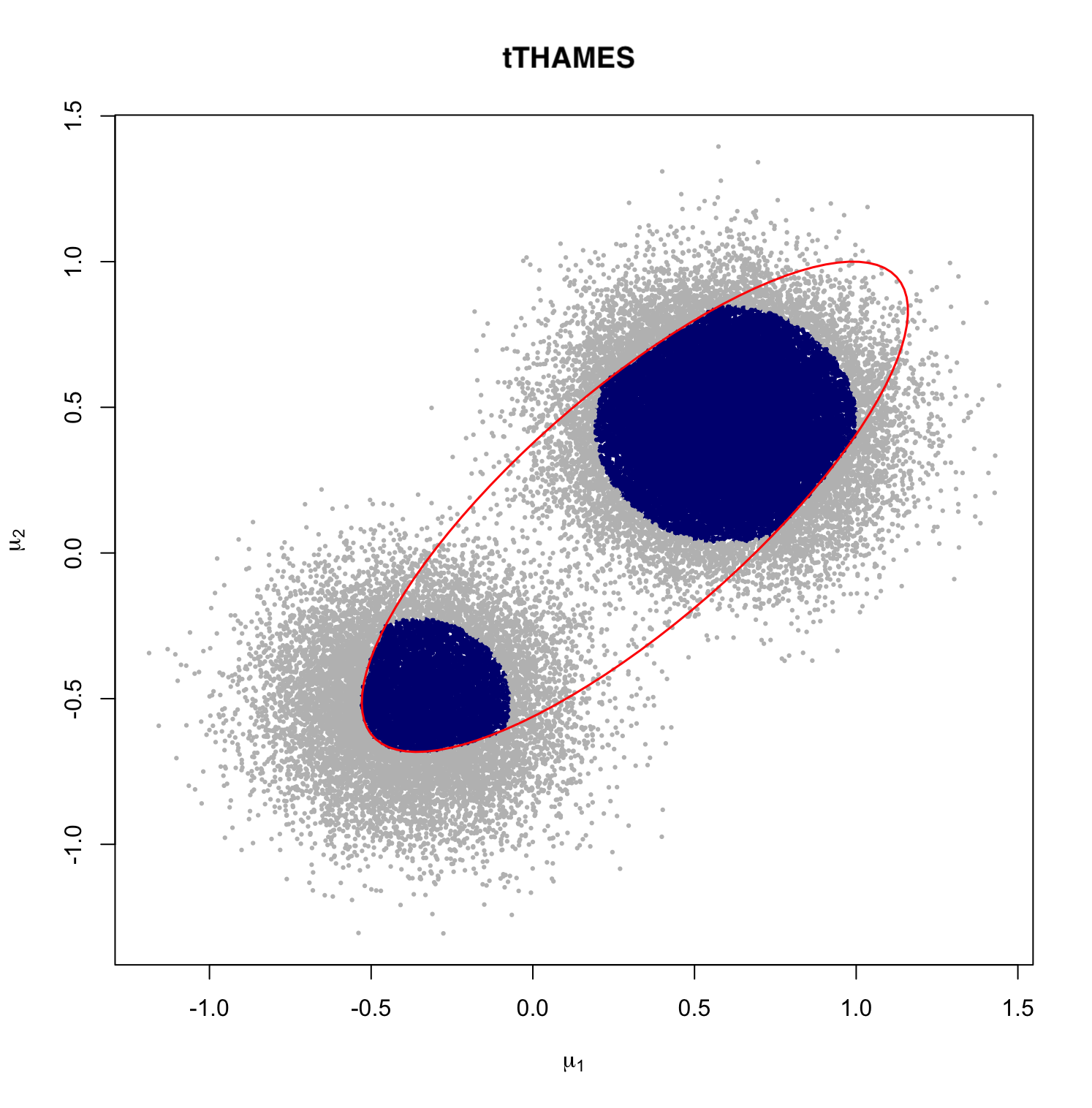}
        \caption{({\bf Example 2}) Shape of sets approximating a HPD region for different methods, when the target is a mixture of two Gaussian posterior distributions. The total number of MCMC simulations is $5 \times 10^4$. Gray dots denote samples drawn from the posterior, while blue dots indicate the posterior sample points used by each estimator.}
    \label{EX2_vis}
\end{figure}

In this multimodal toy example, Figure \ref{EX2_vis} illustrates how the HPD simulations are contributing to the approximate HPD region. In the case of ECMLE, the HPD regions are covered by a union of two ellipsoids and therefore ECMLE is using the target topology. 
Predictably, THAMES fails to account for bimodality and consequently includes some extremely low posterior density regions. The truncated version of THAMES manages to handle this issue since, by construction, the mixture based proposal allows elimination of these low posterior density regions while requiring an independent Monte Carlo evaluation of its volume.

\begin{figure}[h!]
\includegraphics[width=0.9\textwidth]{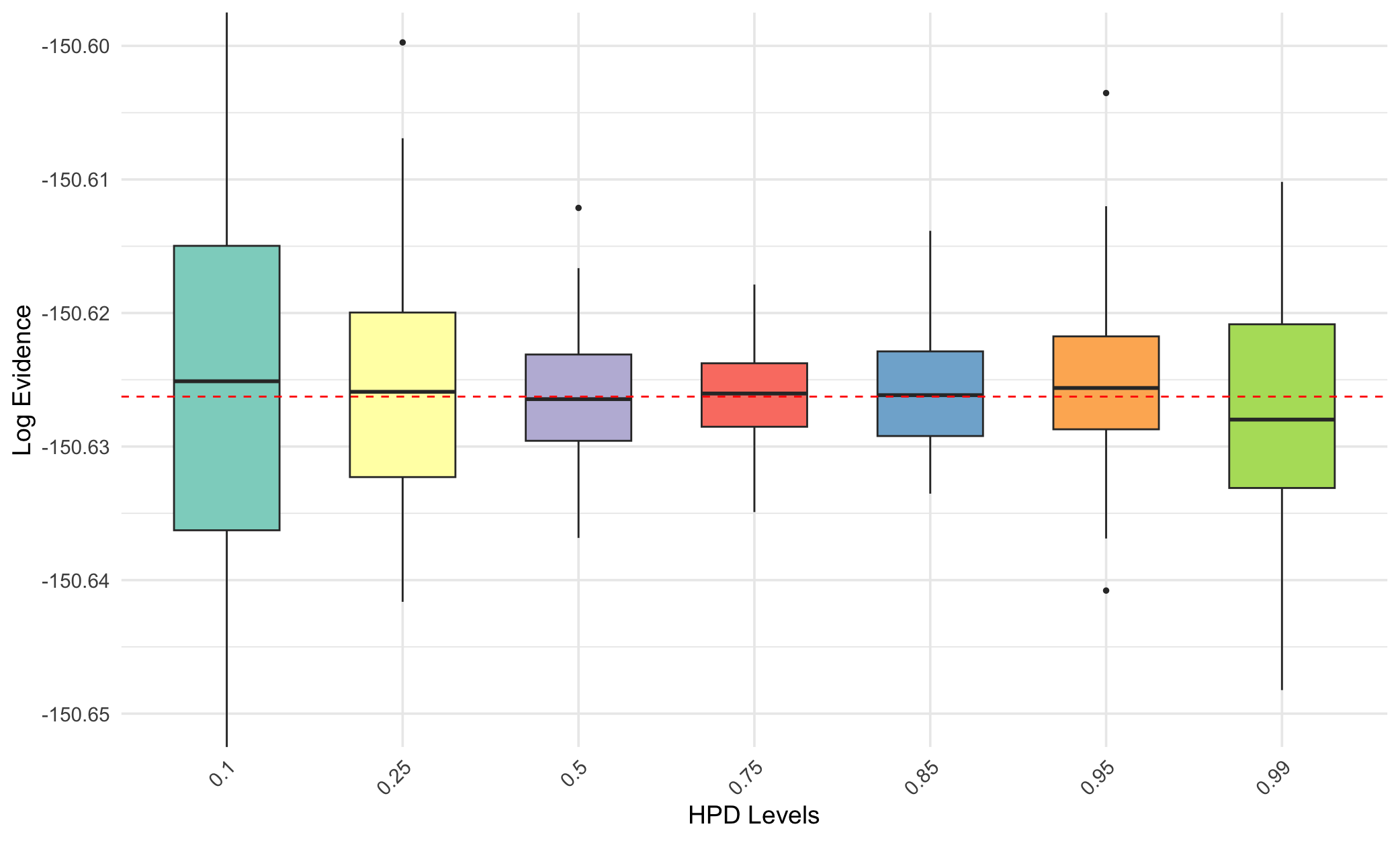}
    \caption{({\bf Example 2}) Comparison of marginal-likelihood estimates across different HPD levels using ECMLE method, for a mixture of two Gaussian posteriors. Each boxplot was created by performing 100 replications of the algorithm, each using $10^5$ posterior draws for the same observed data. The red dashed line indicates the exact log-marginal-likelihood value. 
}
    \label{Ex2_levels}
\end{figure}
Figure~\ref{Ex2_levels} shows the sensitivity of ECMLE to the HPD level $\alpha$ in the bimodal setting, confirming that values around $0.75$ yield the most stable estimates.

\begin{figure}[h!]
\includegraphics[width=0.7\textwidth]{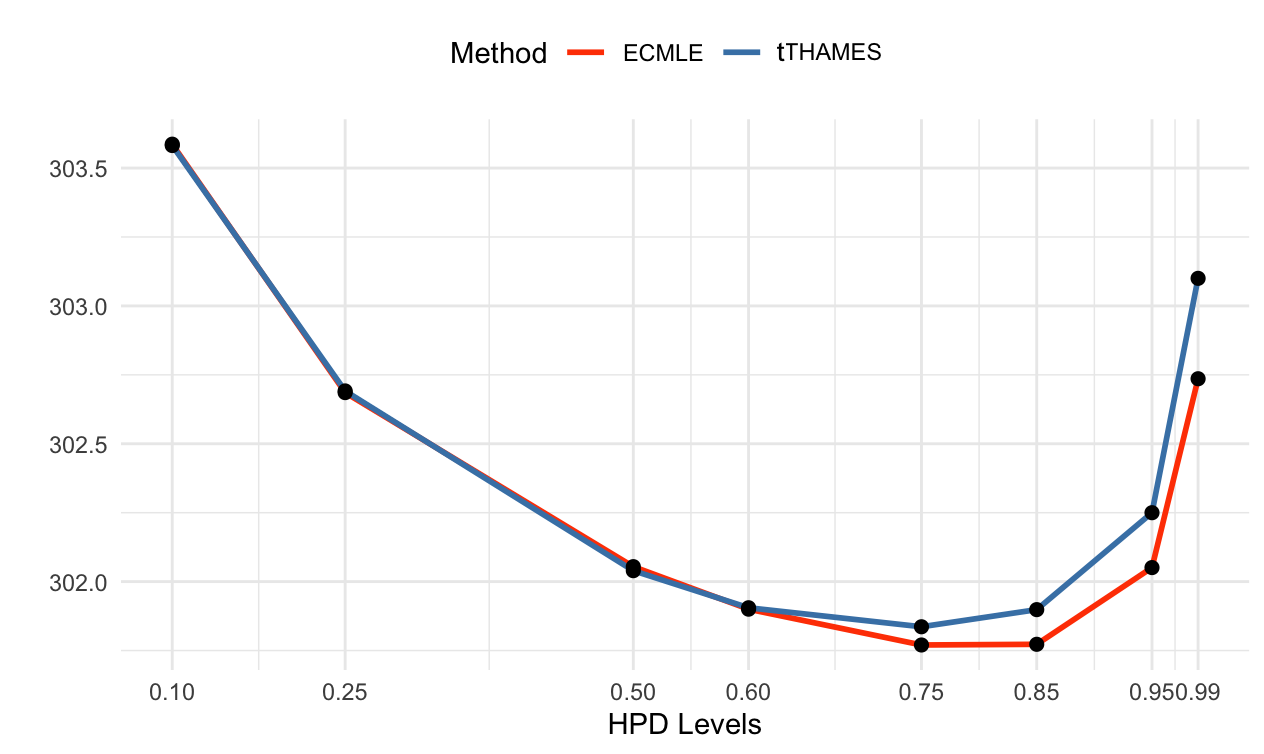}
\caption{({\bf Example 2}) Comparison of the variance 
of the estimators based on the formula \eqref{vareq}, approximated by Monte Carlo for different HPD levels, from $\alpha=0.1$ to $\alpha=0.99$. The number of posterior draws is $10^5$.
}
\label{VR2}
\end{figure}

\begin{figure}[h!]
\includegraphics[width=\textwidth, height=0.3\textheight]{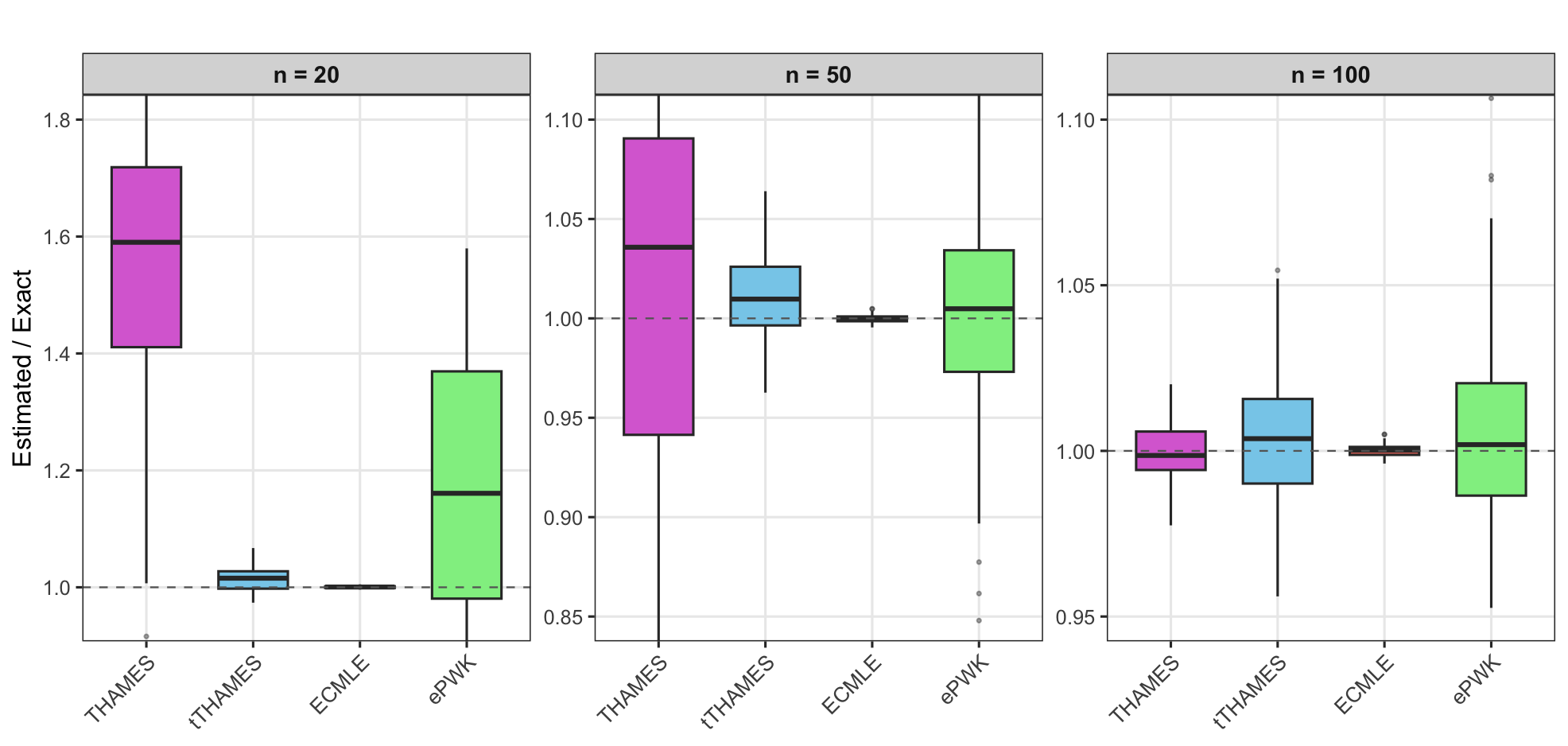}
    \caption{({\bf Example 2}) Boxplots of the evidence ratio ($\widehat{Z}/Z$) for the 2D Gaussian mixture model with $K=2$ and $n \in \{20,50,100\}$, based on $M=100$ datasets. The ideal value is $1$. To ensure a fair comparison at matched runtime, we use per-dataset posterior draws: THAMES (850{,}000), tTHAMES (25{,}500), ECMLE (420{,}000), and ePWK (25{,}000).}

    \label{Ex2_compare}
\end{figure}

To reinforce this initial evaluation, we estimated directly the variance of ECMLE and tTHAMES. Since both estimators are unbiased, the differences between these estimators were contained in the square expectation
\begin{equation}\label{vareq}
\mathbb{E}\left[\hat{Z}^{-2}\right] = \frac{1}{T V(\mathcal{R})^2} \int_{\mathcal{R}} \frac{1}{\pi(\boldsymbol{\theta}) L(\boldsymbol{\theta})} \, \mathrm{d}\boldsymbol{\theta}\,,
\end{equation}
where $\mathcal{R}$ denotes the instrumental region ($\mathcal{E}$ for ECMLE and $A$ for tTHAMES). This quantity can be approximated by a Monte Carlo estimate based on a uniform sample on $\mathcal{R}$. Figure \ref{VR2} confirms the above conclusions and shows that a value of $\alpha$ in the vicinity of $80\%$ achieves better precision than at other HPD levels.
ECMLE also exhibits less variability in estimates compared to the other methods across all sample sizes (n = 20, 50, 100), as shown in Figure \ref{Ex2_compare}, confirming its stability in multimodal settings.

\begin{figure}[h!]
\includegraphics[width=0.3
\textwidth]{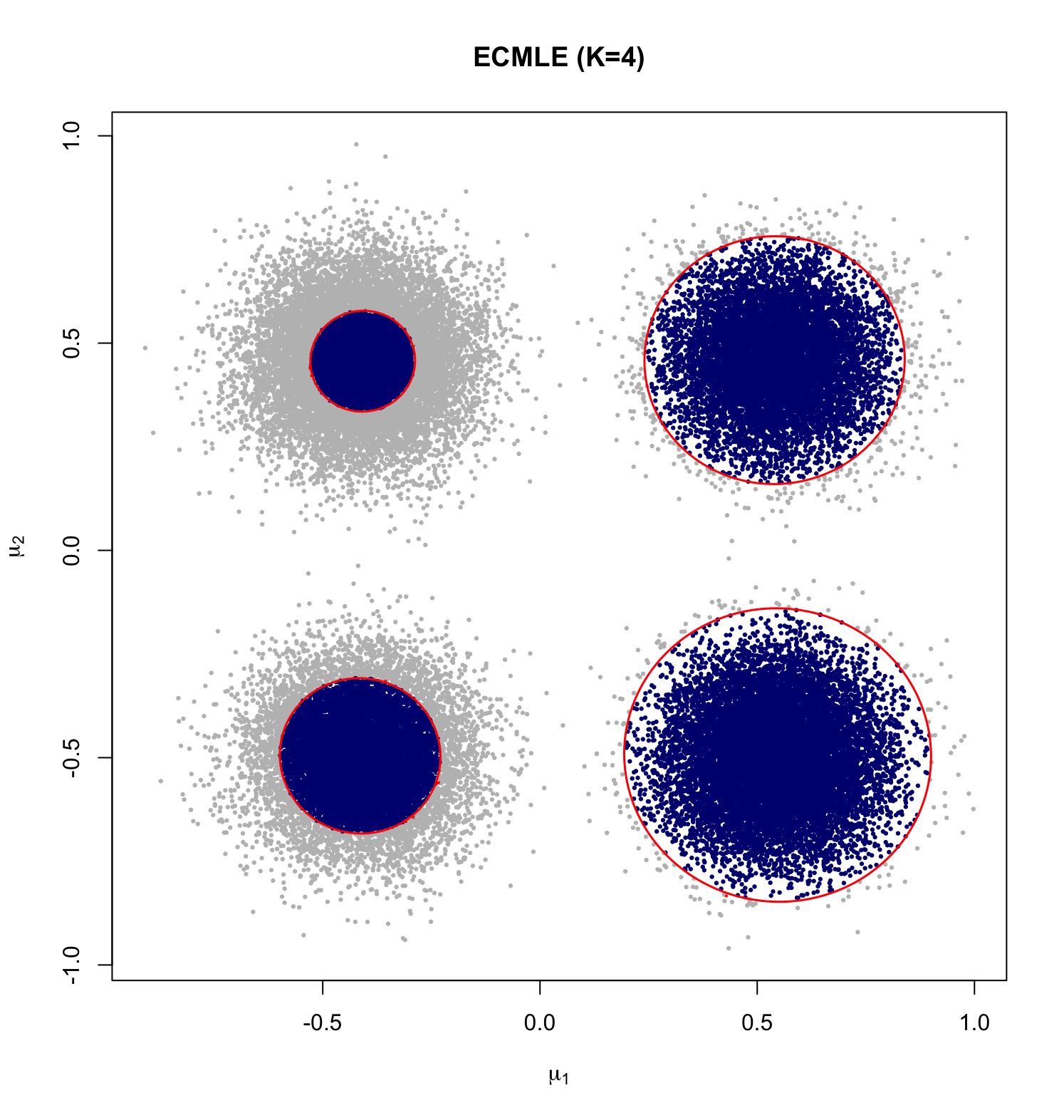}
\includegraphics[width=0.3
\textwidth]{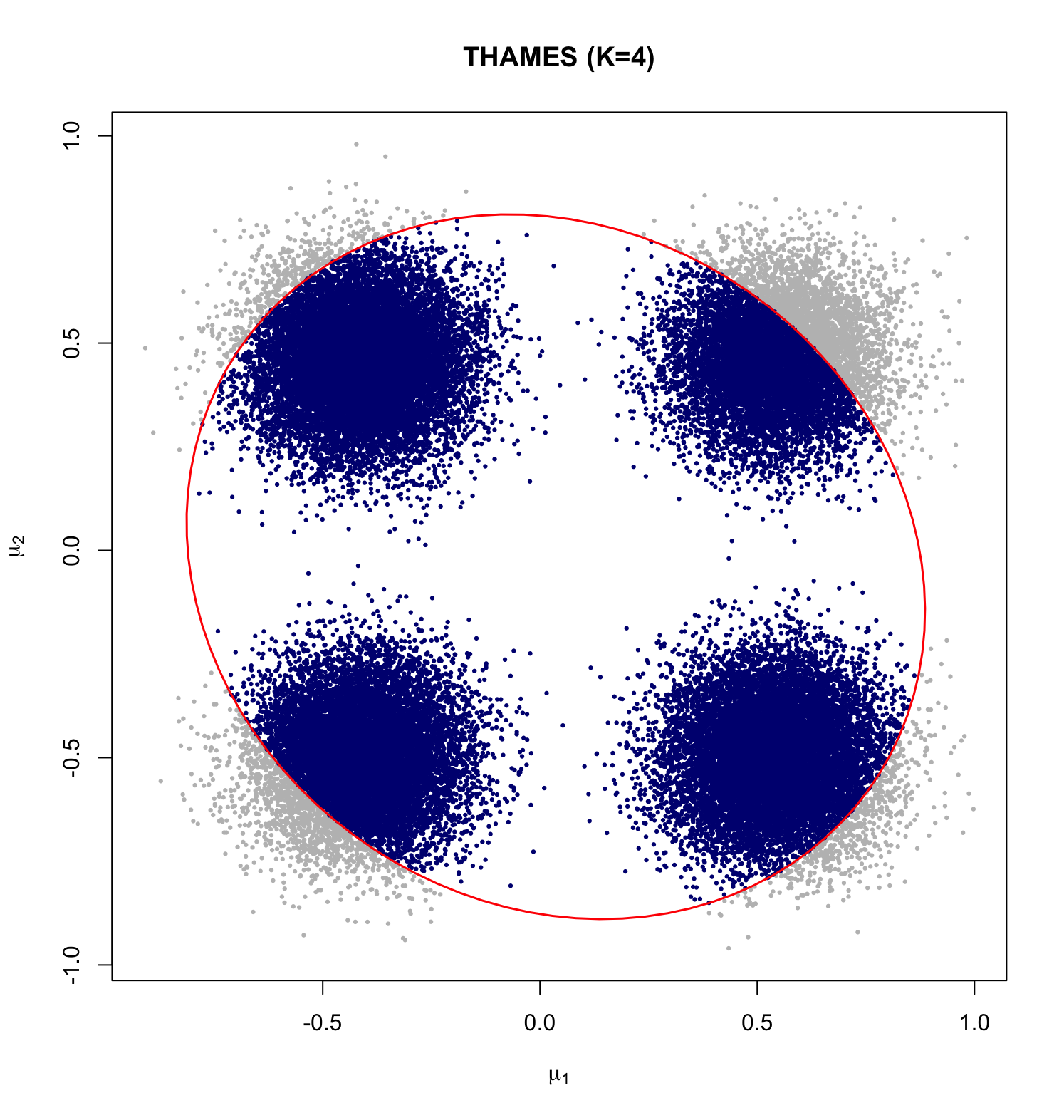}
\includegraphics[width=0.3
\textwidth]{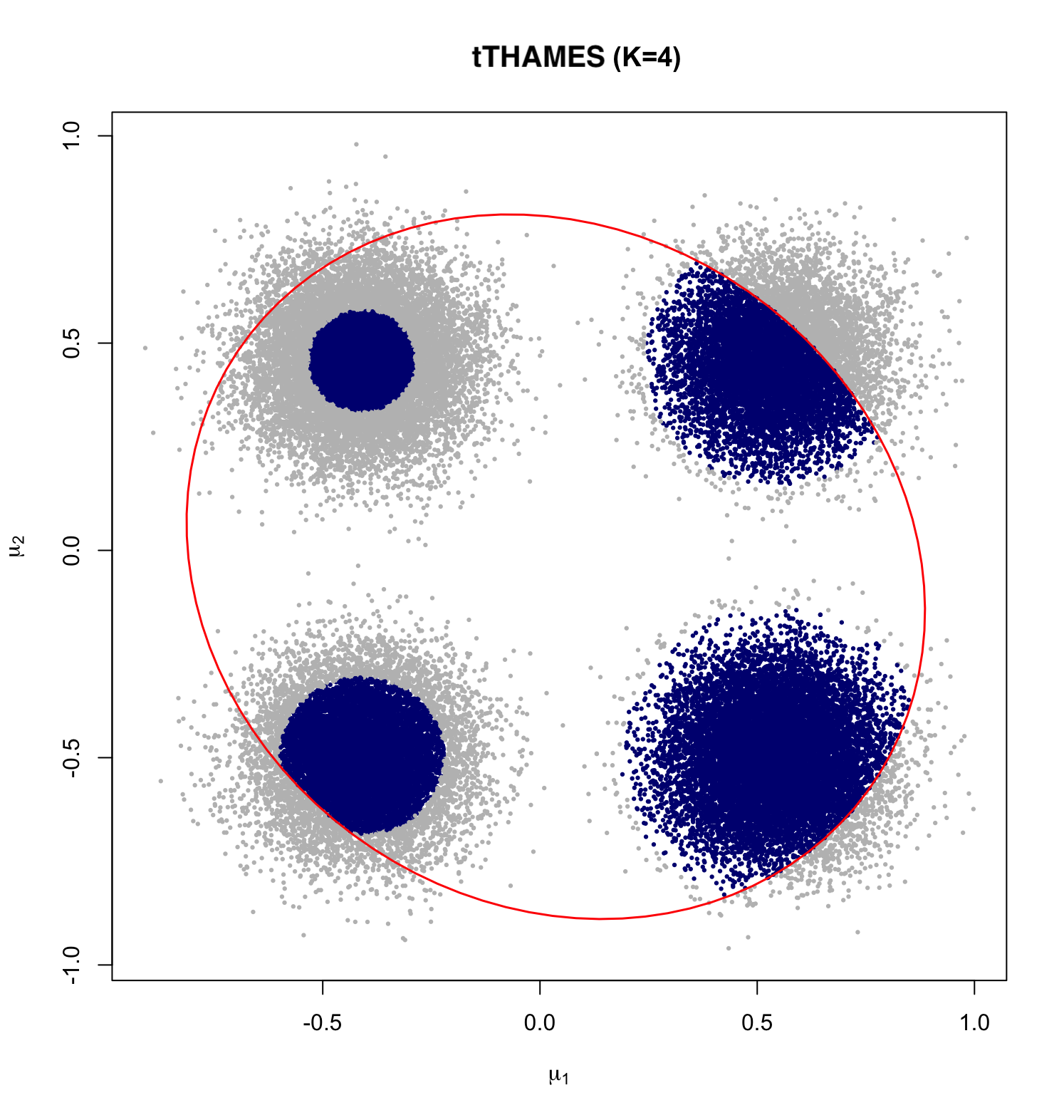}
\\
\includegraphics[width=0.3
\textwidth]{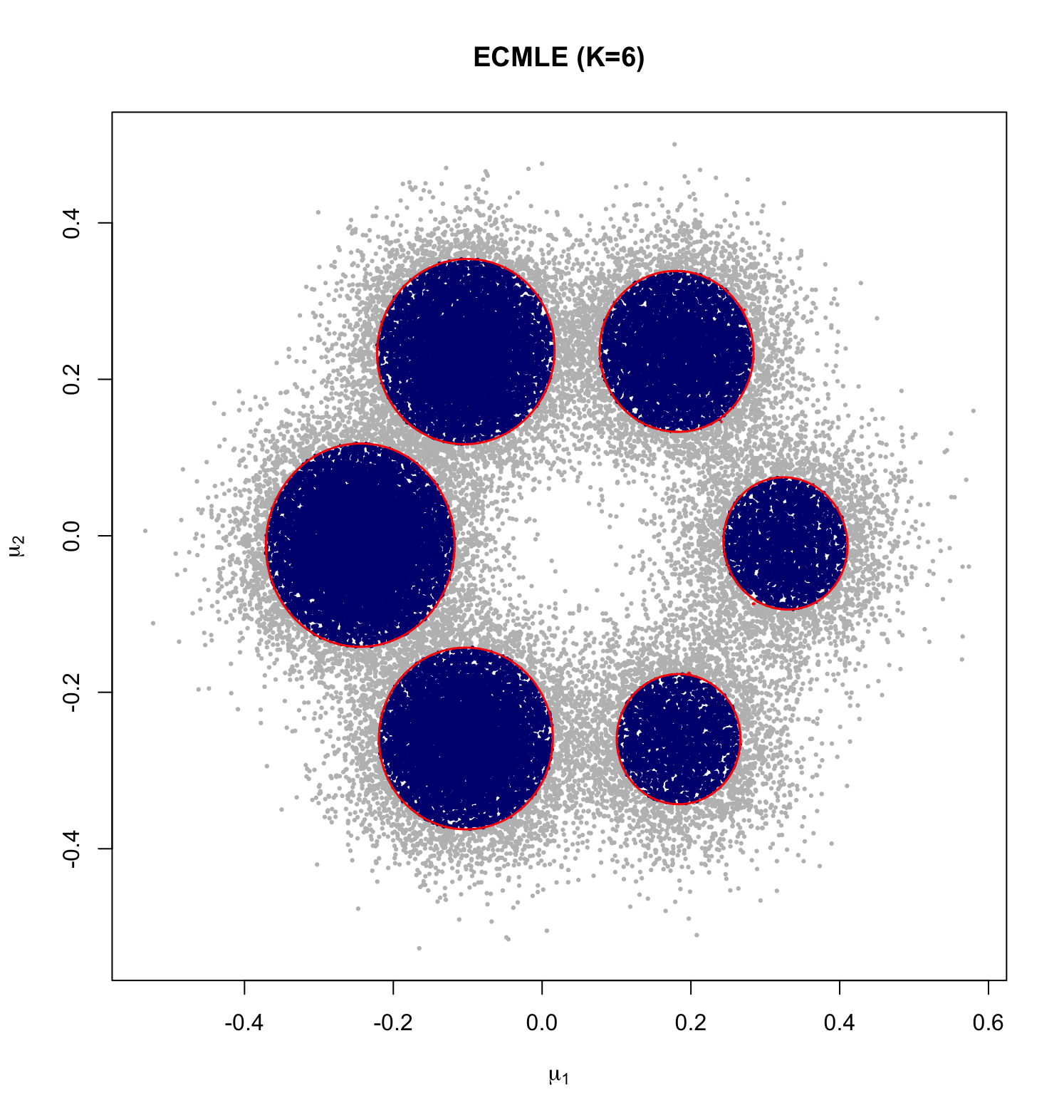}
\includegraphics[width=0.3
\textwidth]{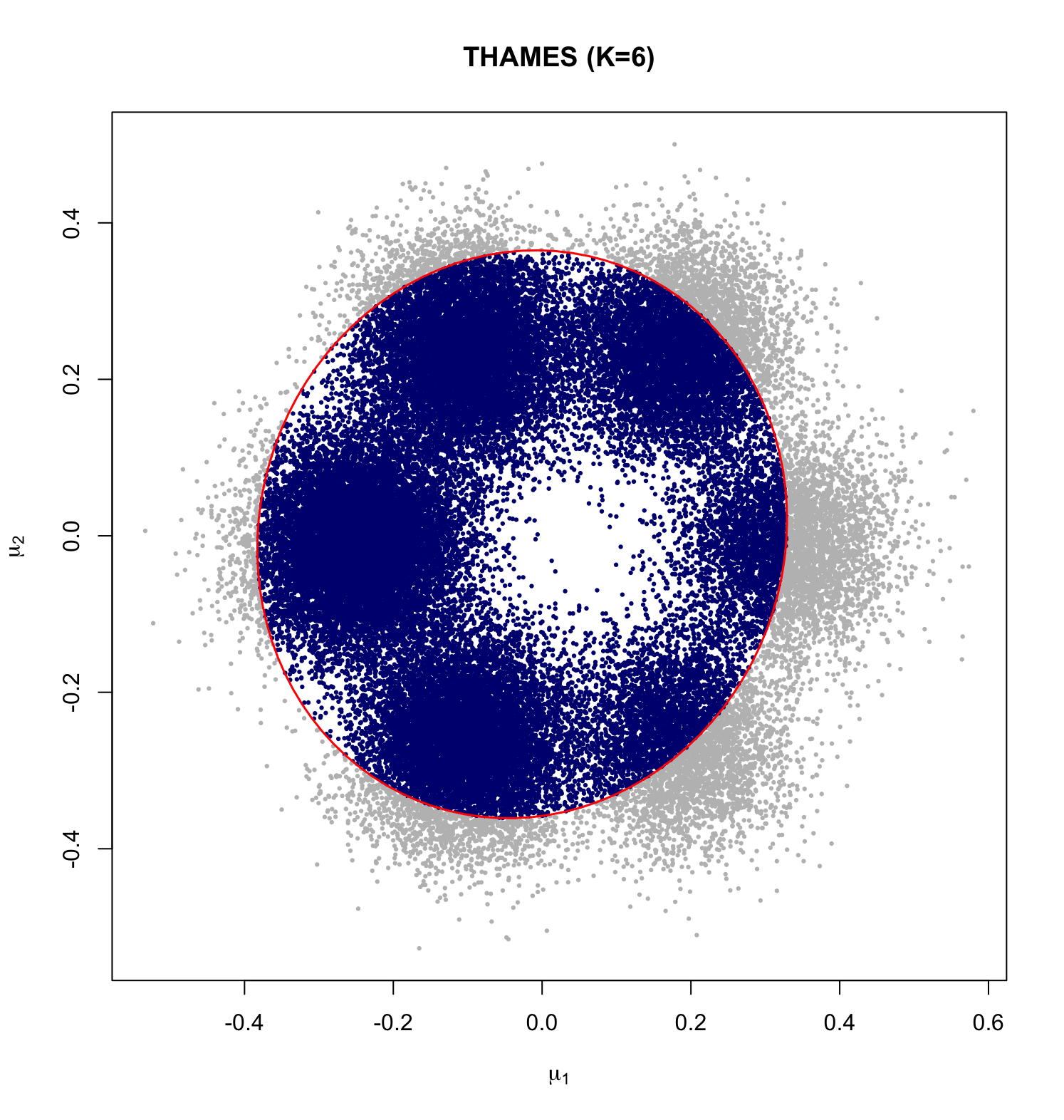}
\includegraphics[width=0.3
\textwidth]{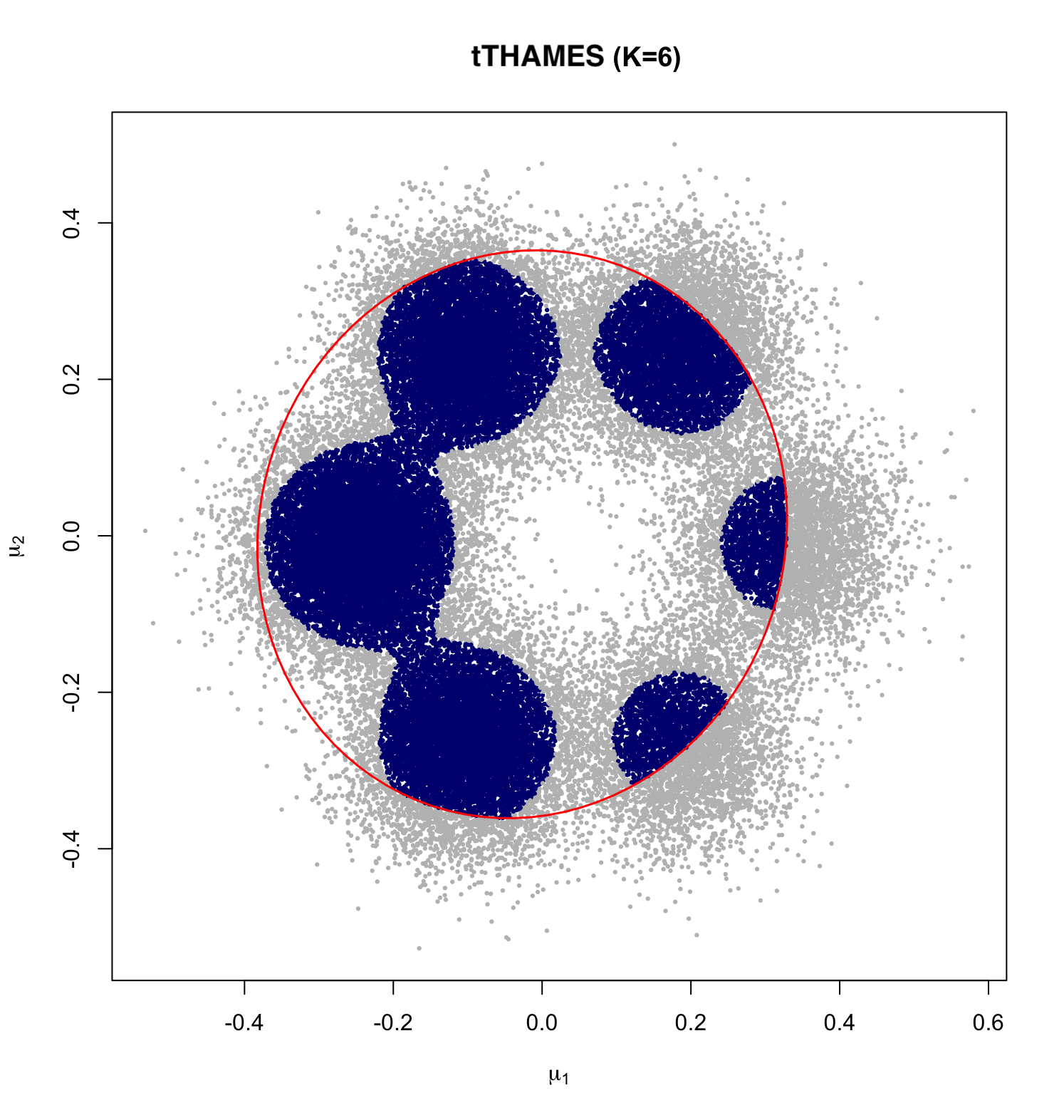}
\\
\includegraphics[width=0.3
\textwidth]{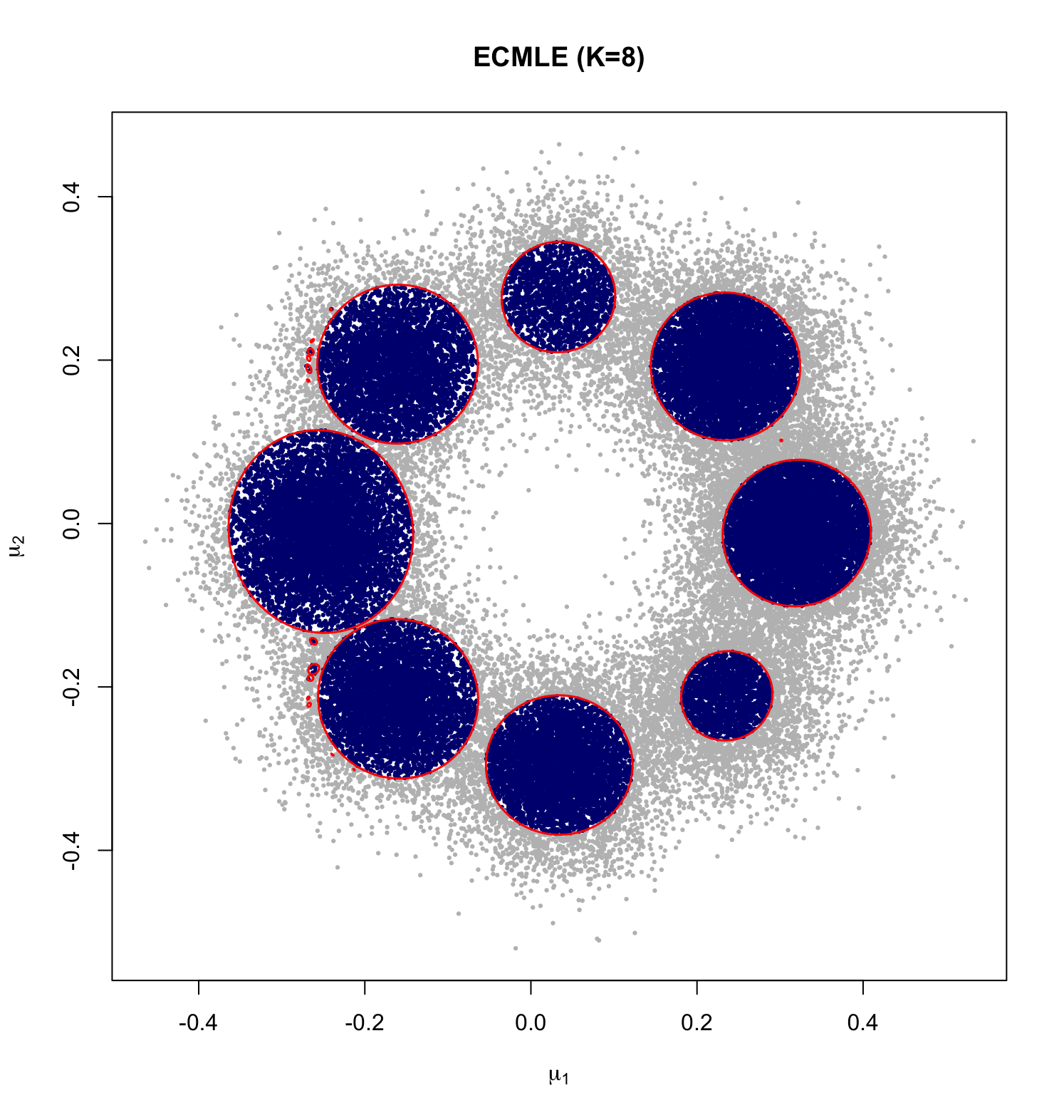}
\includegraphics[width=0.3
\textwidth]{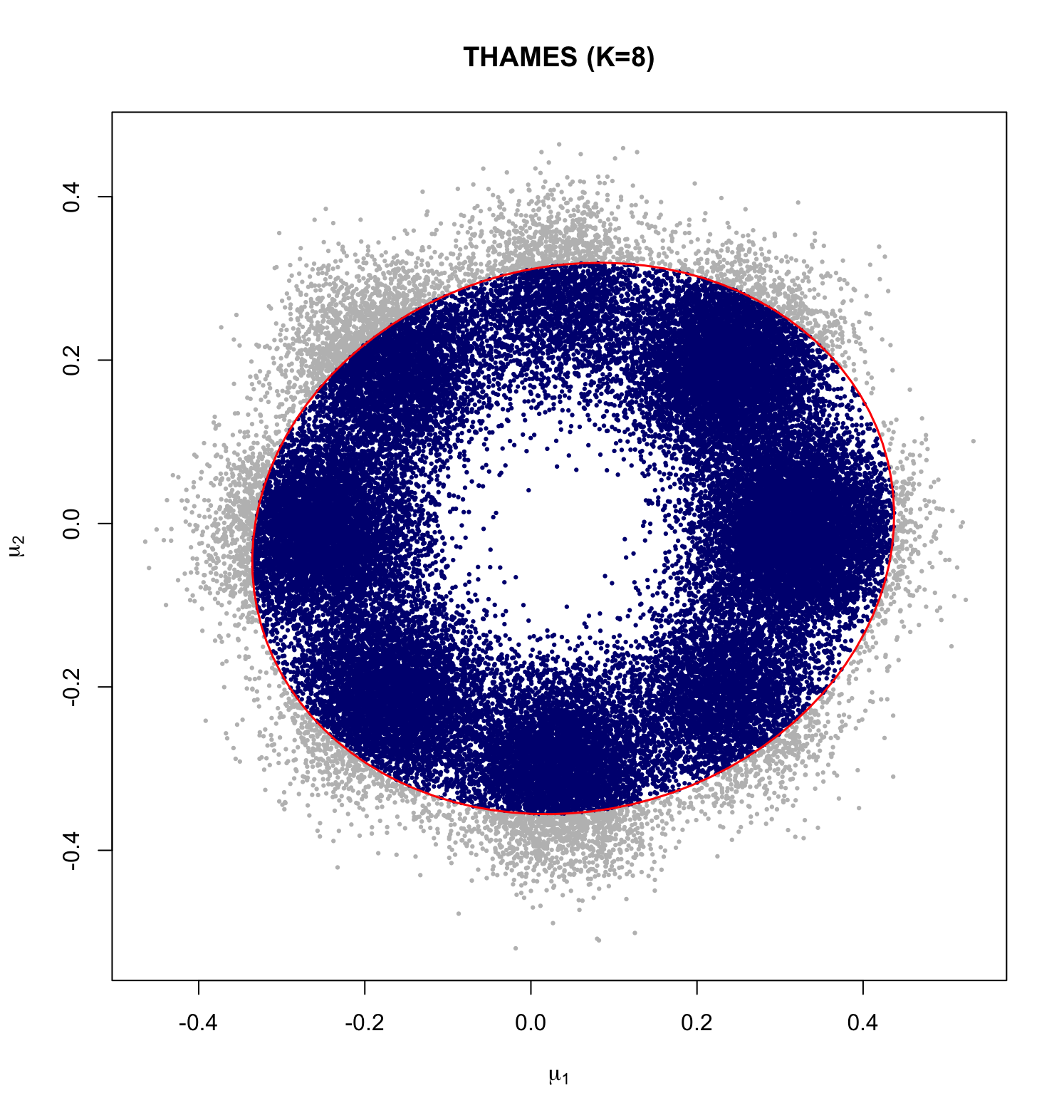}
\includegraphics[width=0.3
\textwidth]{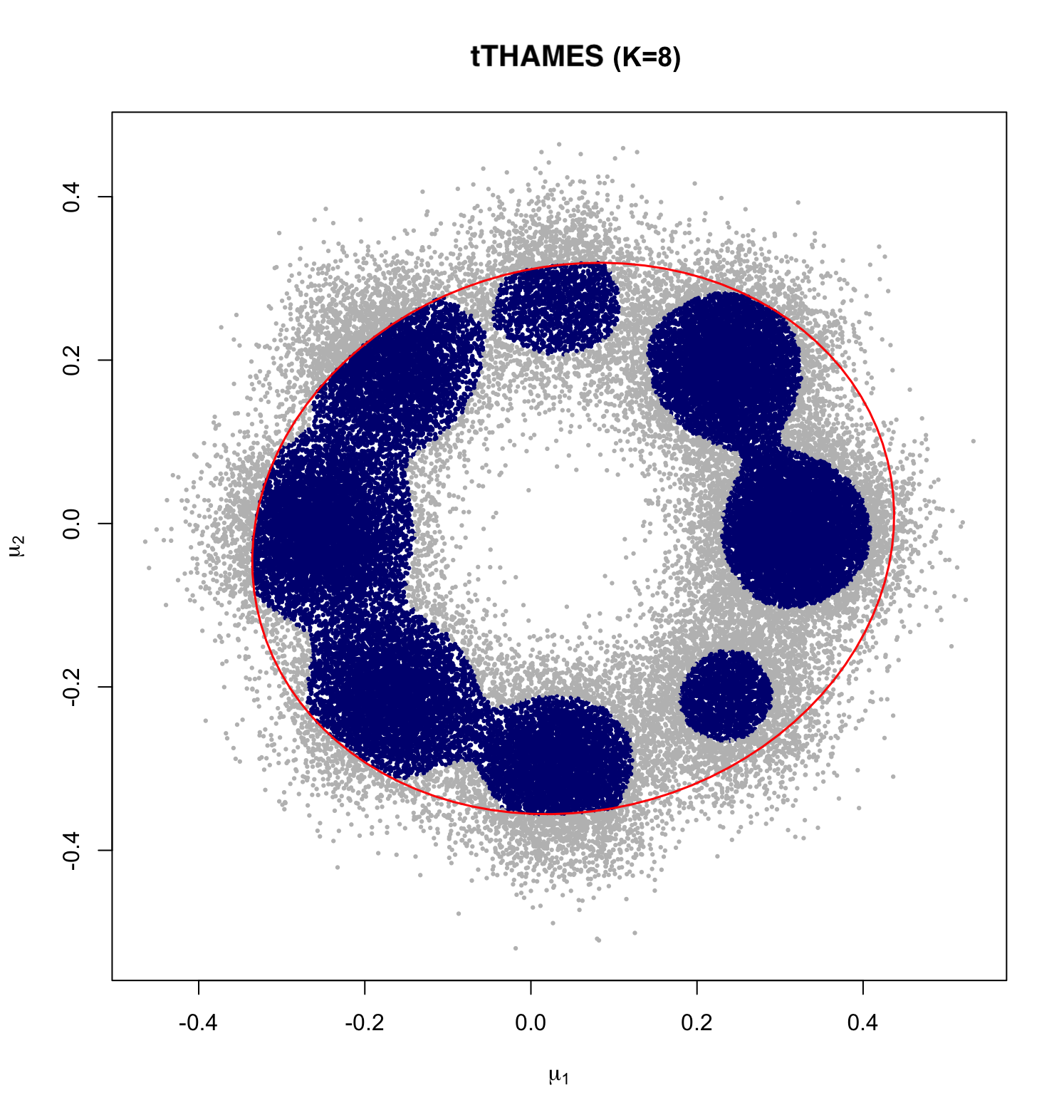}
\caption{({\bf Example 2}) Coverage of the posterior samples: ({\em Left})  ECMLE; ({\em Middle}) THAMES; ({\em Right}) tTHAMES; for three Gaussian mixture posterior distributions with $K=4$, $K=6$, and $K=8$ components. The blue points represent posterior draws considered by each method for estimating the marginal likelihood, and gray points fall outside when the HPD level is 75\%~HPD region. Each panel corresponds to a dataset of size $n=20$ observations and $T=5\times 10^4$ iid simulations from the posterior. All methods are based on the same MCMC simulations and dataset.}
\label{EX2_ellipsoids_created}
\end{figure}

Extending to priors with $K \in \{4, 6, 8\}$ Gaussian mixture components, with observations drawn from $\mathcal{N}(0, \Sigma)$, Figure~\ref{EX2_ellipsoids_created} shows the coverage of the posterior samples using ECMLE, tTHAMES, and THAMES. Each panel corresponds to a single dataset of $n=20$ observations.
For $\alpha=0.75$, Figure~\ref{EX2_ellipsoids_created}, ECMLE with $K=4$ (top left), shows four red ellipsoids indicating the $75\%$ high-density region that has been perfectly covered by ECMLE. The blue points represent posterior samples within the $\alpha$-HPD that fall inside the ECMLE ellipsoids, while gray ones fall outside these regions.
Figure~\ref{EX2_ellipsoids_created}, THAMES and tTHAMES (middle and right columns), display their ellipsoids by red lines. For tTHAMES (right column), among the points within the ellipsoid, blue points correspond to those with the $\alpha$ highest posterior density values. By comparing with ECMLE (left column), we observe that some of the $\alpha$-HPD points have not been covered by tTHAMES. 
Furthermore, for THAMES (middle column), a significant portion of non-HPD posterior samples fall within the ellipsoid, highlighting cases where a single ellipsoid fails to adequately capture the high-density region. For $K=6$, ECMLE covered $71.82\%$ of the $\alpha$-HPD region, tTHAMES covered $67.83\%$, and the THAMES ellipsoid included $82.51\%$ of all posterior samples.

Moreover, as shown by Figure \ref{EX2_ellipsoids_created}, ECMLE achieves accurate HPD approximation through its adaptive boundary-aware design. Unlike fixed geometric shapes, ECMLE places ellipsoids exclusively at HPD sample locations and calculates each semi-axis to ensure maximal coverage within high-density regions while precisely respecting the HPD boundaries. This adaptive semi-axes calculation allows the method to conform locally to arbitrary posterior geometries (whether multimodal, skewed, or irregularly shaped) without the geometric constraints of THAMES that may systematically include low-density areas or miss complex boundary structures.

\begin{figure}[h!]
\includegraphics[width=\textwidth]{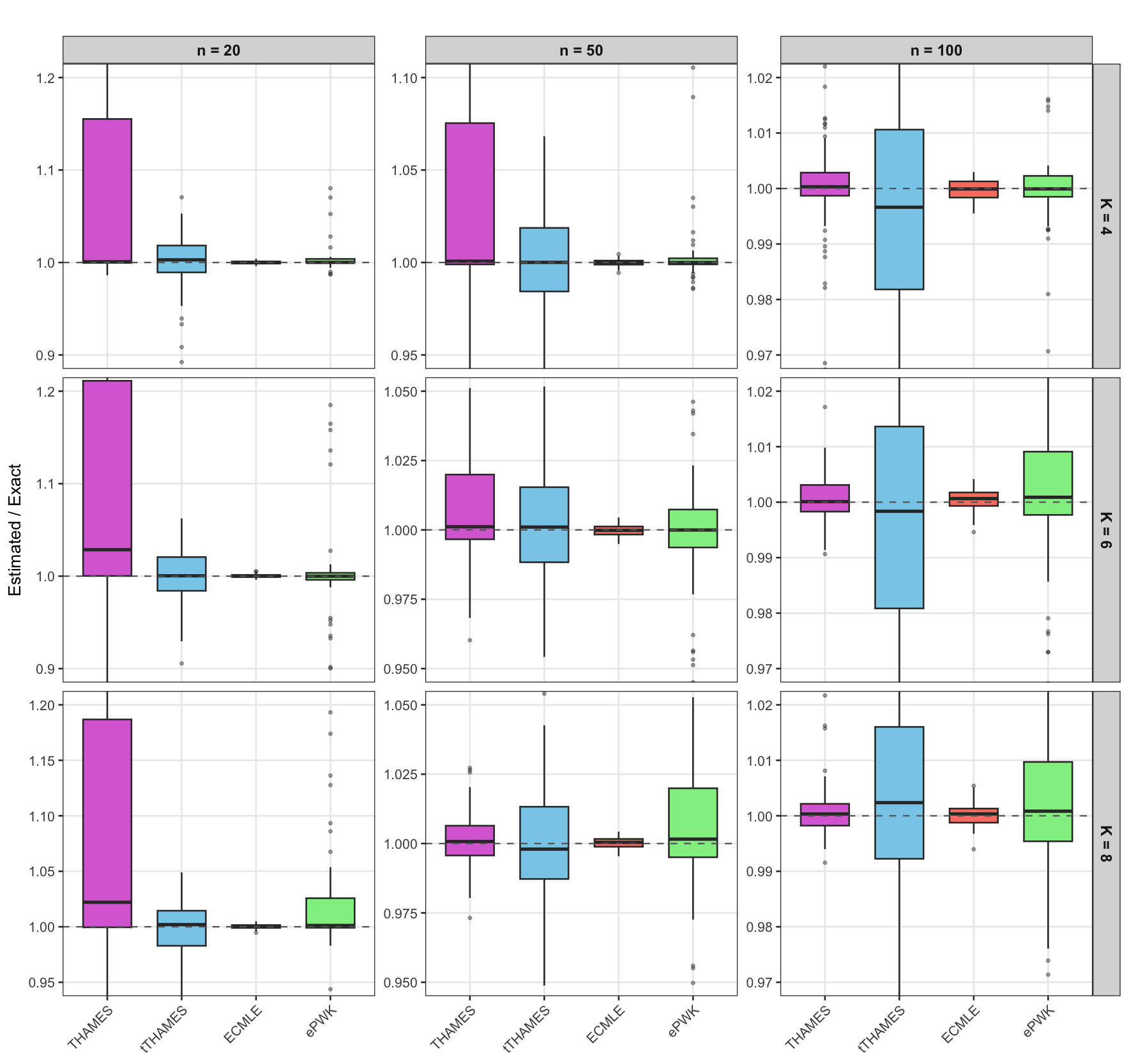}
    \caption{({\bf Example 2}) Comparison of evidence ratio ($\widehat{Z}/Z$) for THAMES, tTHAMES, ECMLE, and ePWK across nine configurations of the Gaussian mixture model: $K \in \{4,6,8\}$ mixture components and sample sizes $n \in \{20, 50, 100\}$. All methods were calibrated so that each runs under approximately the same computational budget, ensuring that comparisons reflect estimation accuracy rather than differences in runtime. }
    \label{multimode}
\end{figure}

To further evaluate the performance of the estimators across varying model complexity and sample sizes, Figure \ref{multimode} presents a comprehensive comparison of the evidence ratio ($\widehat{Z}/Z$) for THAMES, tTHAMES, ECMLE, and ePWK across nine configurations of the Gaussian mixture model, with $K \in \{4, 6, 8\}$ mixture components and sample sizes $n \in \{20, 50, 100\}$. All methods were calibrated to operate under approximately the same computational budget, ensuring a fair comparison based on estimation accuracy rather than runtime differences. Across all configurations, ECMLE consistently achieves estimates closest to the true evidence value with the lowest variability. As the number of mixture components increases, the advantage of ECMLE becomes more pronounced, reflecting its ability to adapt to increasingly complex multimodal posterior geometries. In contrast, THAMES exhibits substantial bias and high variance, particularly for larger $K$, as its single ellipsoid fails to capture the multiple modes of the posterior. The tTHAMES improves upon THAMES but still shows greater variability than ECMLE, especially for smaller sample sizes. The ePWK estimator, while competitive in lower-dimensional symmetric cases, struggles as the posterior complexity grows, highlighting its sensitivity to deviations from ideal geometric conditions.

\subsubsection*{Example 3: Rosenbrock Distribution}
In this example, we examine the performances of the different methods previously mentioned for a posterior distribution with a boomerang shape called the {\em Rosenbrock distribution} that has often been used as a benchmark in the Bayesian computational literature (\cite{haario1999adaptive,wraith2009estimation, pagani2022}).

More precisely, each entry $X_i \in \mathbb{R}^{d-1}$ ($i=1, \dots, n$) follows the density
\[
p(x \mid \pmb{\theta}) \propto \exp\left( -\frac{1}{2\sigma^2} \sum_{j=2}^d (x_{(j-1)} - \theta_j - b_{j-1} \{\theta_{j-1}^2 - a_{j-1}\})^2 \right),
\]
where $\pmb{\theta} = (\theta_1, \dots, \theta_d)^\top \in \mathbb{R}^d$ is the parameter vector, and $a, b \in \mathbb{R}^{d-1}$ are fixed constants. This formulation captures a chain of quadratic dependencies, starting from $\theta_1$ without a direct observation and propagating through subsequent dimensions. The likelihood for $n$ independent observations is then
\[
L(\pmb{\theta}) \propto \exp\left( -\frac{1}{2\sigma^2} \sum_{i=1}^n \sum_{j=2}^d (x_{i,(j-1)} - \theta_j - b_{j-1} \{\theta_{j-1}^2 - a_{j-1}\})^2 \right).
\]

In our specific example, we leverage sufficient statistics to simplify the model. We consider a $d$-dimensional formulation where each parameter $\theta_j$ has a corresponding observation $\bar{Y}_j$, enabling both exact posterior sampling and a closed-form marginal likelihood. Let $\bar{Y} = (\bar{Y}_1, \bar{Y}_2, \dots, \bar{Y}_d)^\top \in \mathbb{R}^d$ denote the vector of sample means, where $\bar{Y}_j = \frac{1}{n} \sum_{i=1}^n x_{i,j}$ for $j=1, \dots, d$. Here $\bar{Y}_1$ directly informs $\theta_1$, while the remaining observations $\bar{Y}_2, \ldots, \bar{Y}_d$ retain the Rosenbrock nonlinear dependence on preceding parameters.

The conditional distributions are
\[
\bar{Y}_j \mid \pmb{\theta} \sim \mathcal{N}(\mu_j(\pmb{\theta}), \sigma^2 / n), \quad j = 1, \dots, d,
\]
with
\[
\mu_1(\pmb{\theta}) = \theta_1, \quad \mu_j(\pmb{\theta}) = \theta_j + b_{j-1} (\theta_{j-1}^2 - a_{j-1}) \quad \text{for } j=2, \dots, d.
\]

This structure maintains the Rosenbrock dependencies while incorporating observations across all dimensions. The full likelihood density becomes
\[
p(\bar{Y} \mid \pmb{\theta}) = \left( 2\pi \frac{\sigma^2}{n} \right)^{-d/2} \exp\left( -\frac{n}{2\sigma^2} \sum_{j=1}^d (\bar{Y}_j - \mu_j(\pmb{\theta}))^2 \right).
\]

Here, we adopt improper flat priors on all parameters: $\pi(\pmb{\theta}) \propto 1$. Although improper priors may lead to improper posteriors, here the posterior is proper since each parameter $\theta_j$ is directly informed by its corresponding observation $\bar{Y}_j$.

The marginal likelihood $Z$ is obtained by integrating the likelihood over the prior:
$$Z = \int p(\bar{Y} \mid \pmb{\theta}) \, \pi(\pmb{\theta}) \, \text d\pmb{\theta} = 1$$
(see supplementary material for the detailed calculation). This constant evidence highlights a key property of the model under improper priors: the marginal likelihood is data-independent, which can be useful for certain theoretical analyses but may not hold in more constrained settings.

\begin{figure}[h!]
\includegraphics[width=0.3\textwidth]{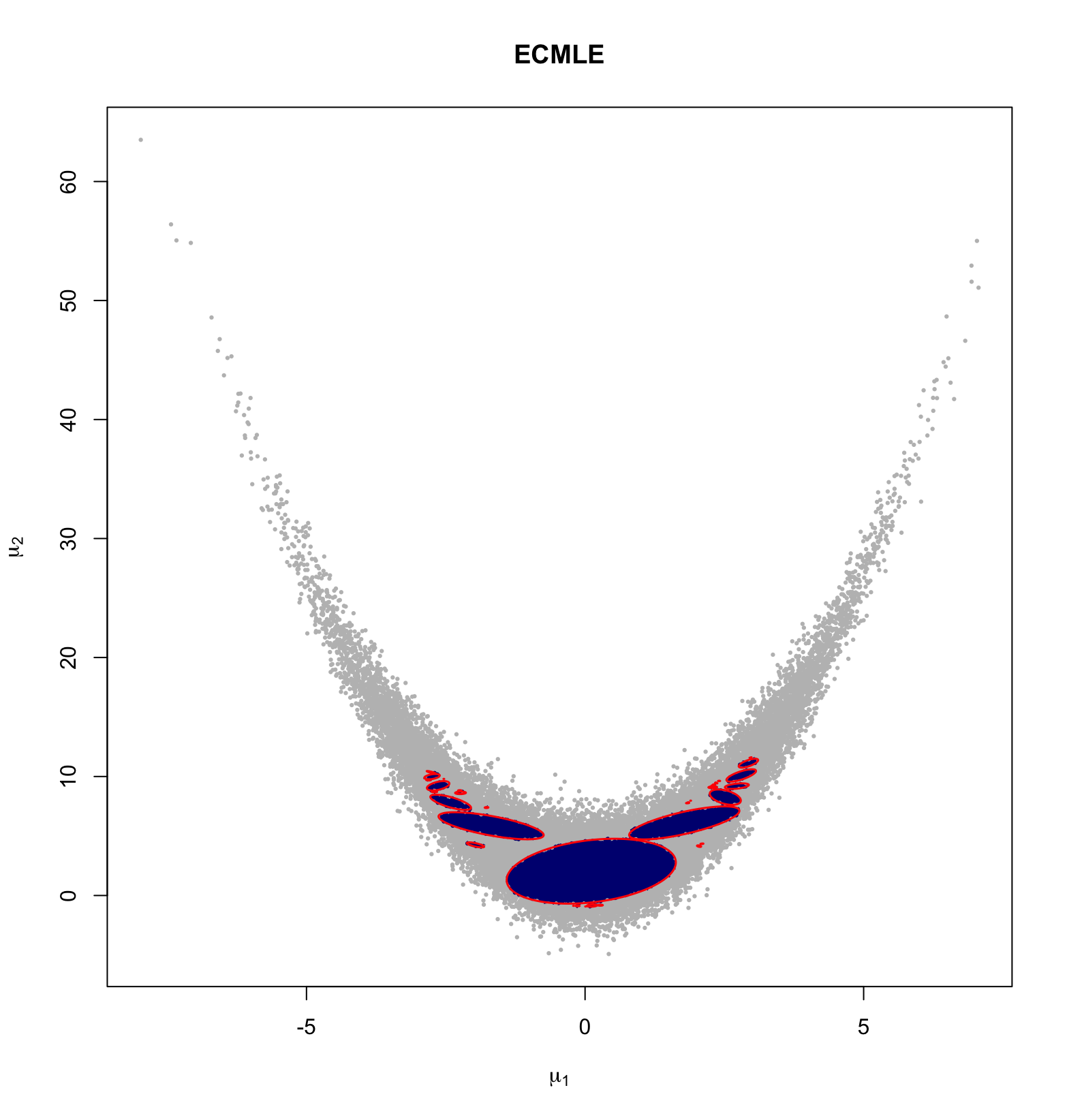}
\includegraphics[width=0.314\textwidth]{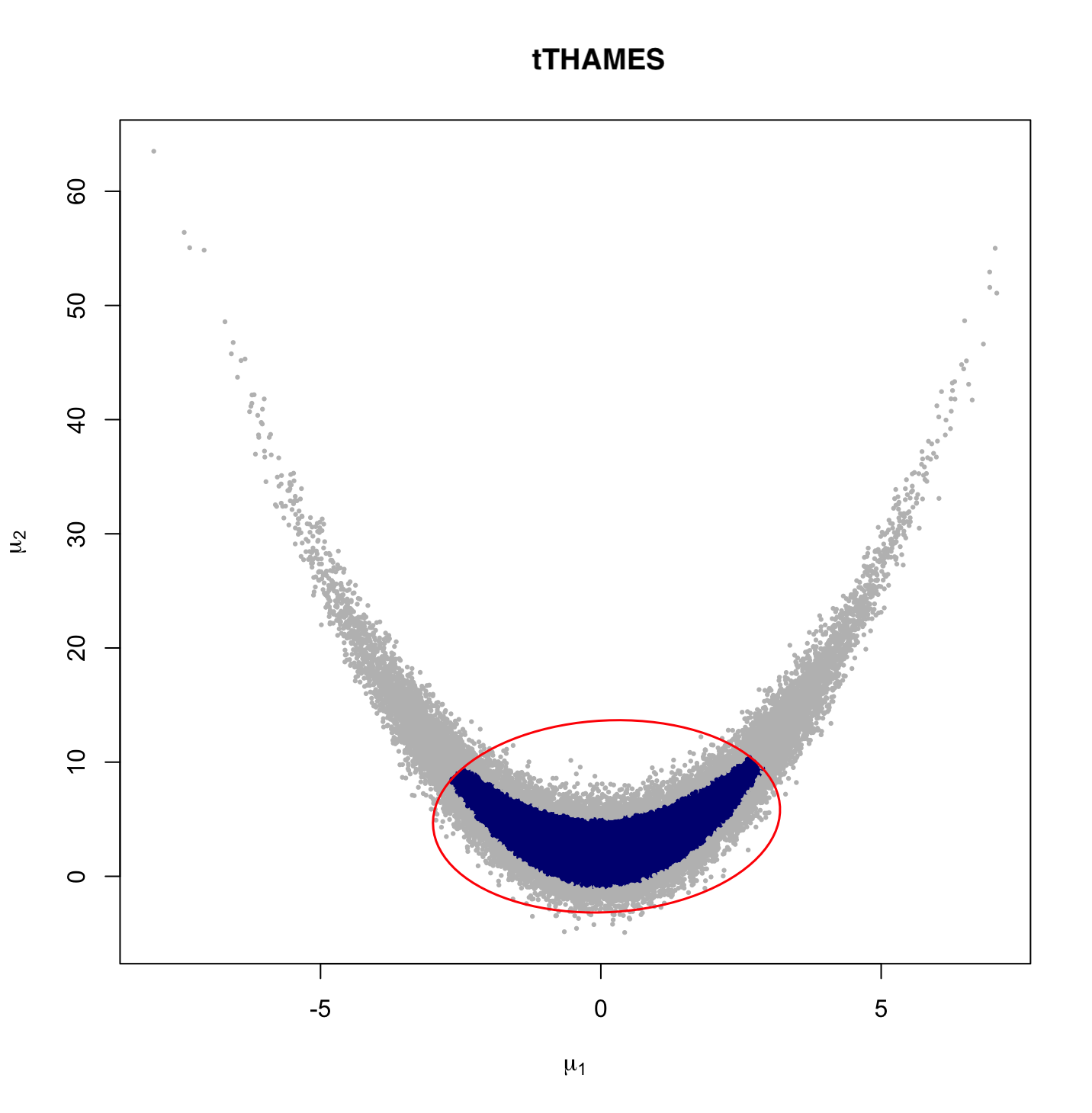}
\includegraphics[width=0.3\textwidth]{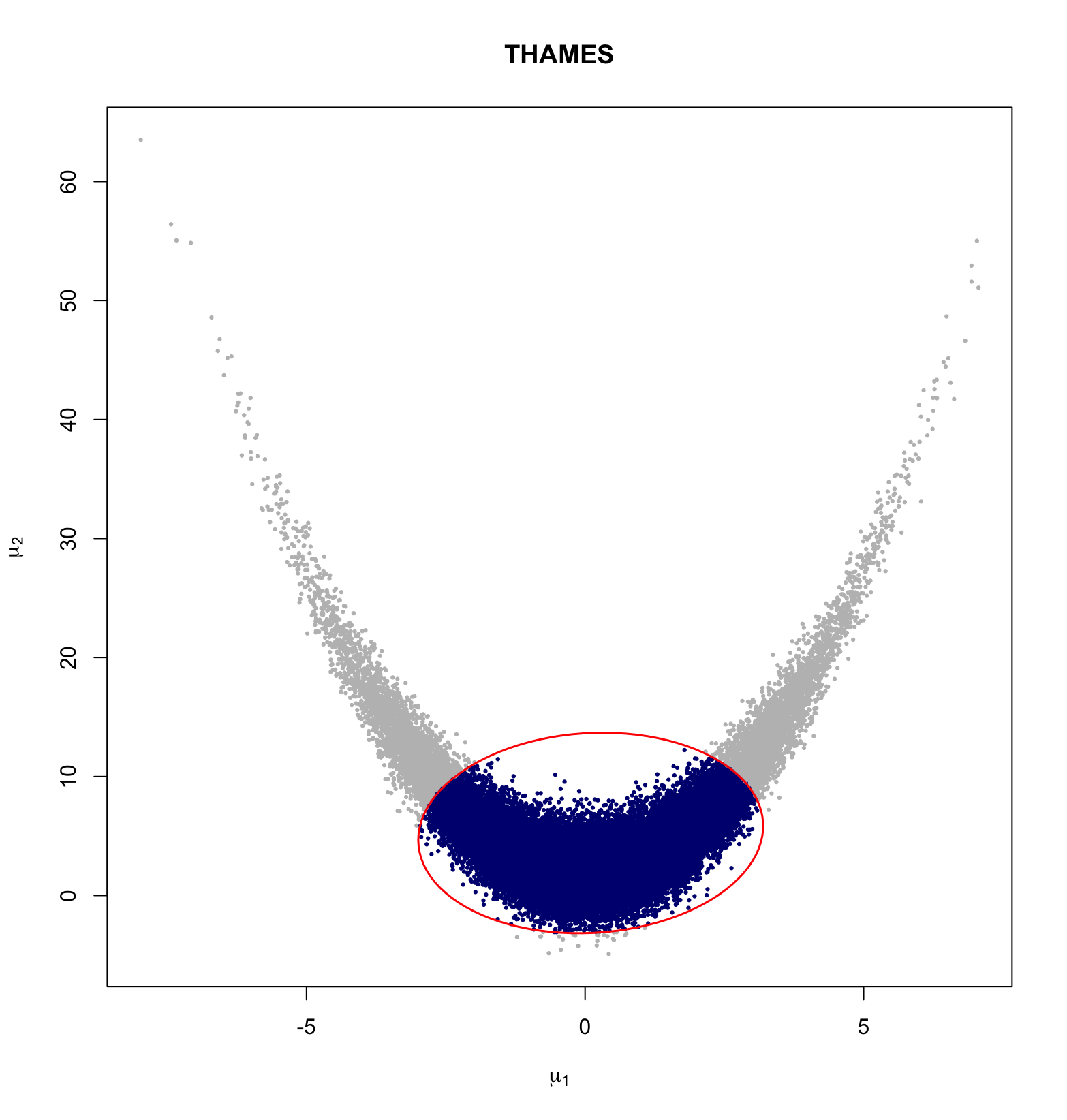}
        \caption{({\bf Example 3}) Approximations of the $\alpha=0.75$ HPD region for the two-dimensional Rosenbrock posterior distribution. 
        }
    \label{fig:Rosen_HPD}
\end{figure}

Figure \ref{fig:Rosen_HPD} illustrates the different coverages of the $\alpha$-HPD region by the three methods and highlights the difficulties of THAMES in fitting the non-linear structure of the posterior surface. In such a case, since the covariance matrix of the observations does not convey useful information about the shape of the posterior, tTHAMES is also missing a part of the HPD region. In contrast, ECMLE extends further and better represents the full structure of the HPD region, including its separate branches.

\begin{figure}[h!]
\includegraphics[width=\textwidth]{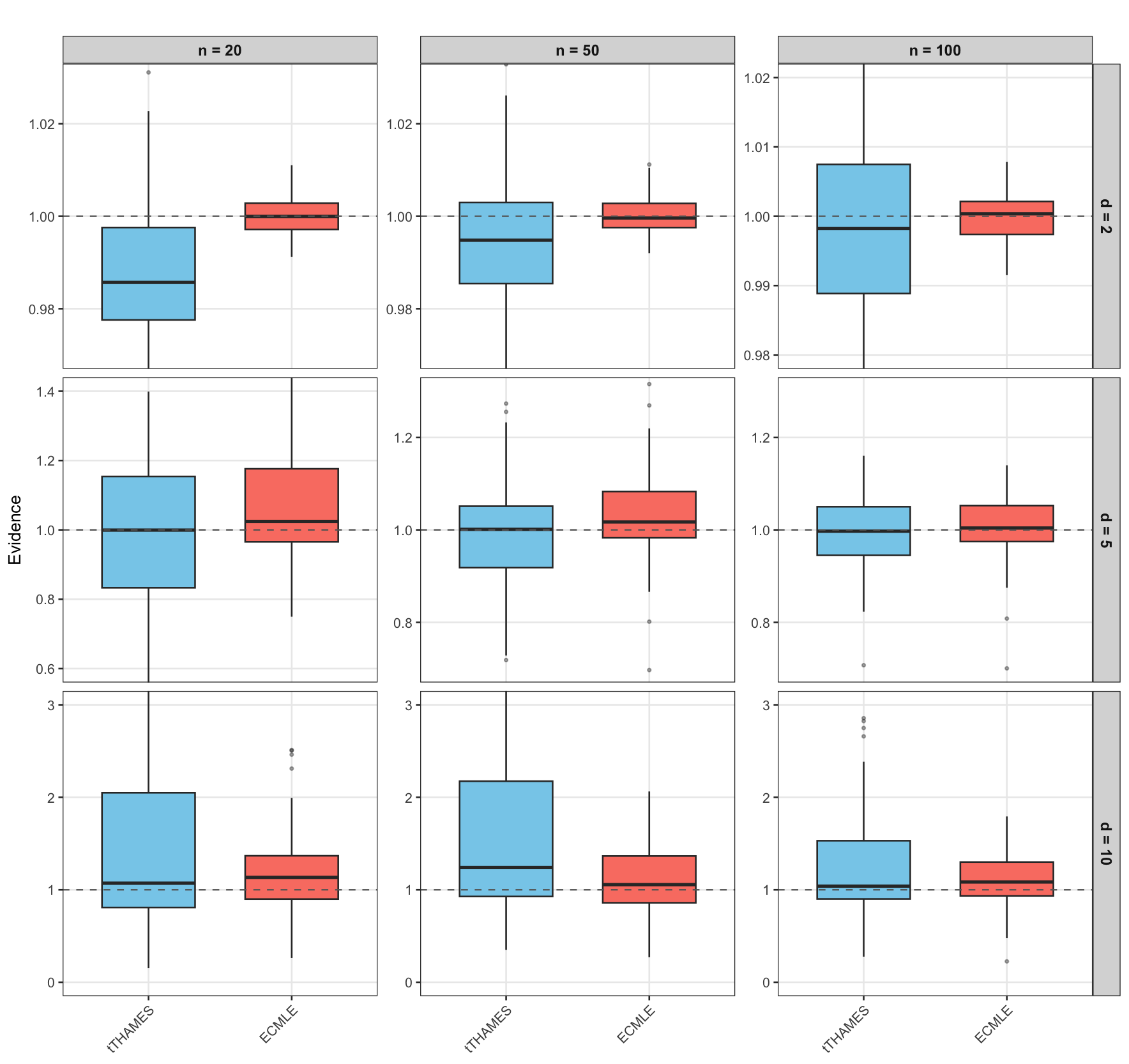}
        \caption{(\textbf{Example 3}) Boxplots of marginal likelihood estimates for the Rosenbrock distribution posterior in dimensions $d \in \{2,5,10\}$ and sample sizes $n \in \{20,50,100\}$, based on $M=100$ independently generated datasets. For each dataset, the remaining methods use the same posterior sample, but with method-specific numbers of draws chosen so that all methods operate under a comparable computational budget. For $d=2$: tTHAMES (25{,}500), ECMLE (420{,}000). For $d=5$: tTHAMES (48{,}000), ECMLE (350{,}000). For $d=10$: tTHAMES (67{,}000), ECMLE (28{,}000). The dashed horizontal line marks the exact marginal likelihood ($Z=1$).}
    \label{fig:Rosenall}
\end{figure}

Figure \ref{fig:Rosenall} presents a comprehensive evaluation of marginal likelihood estimates for the Rosenbrock distribution across dimensions $d \in \{2, 5, 10\}$ and sample sizes $n \in \{20, 50, 100\}$, based on 100 independently generated datasets. Due to their extremely high variances in this challenging setting, THAMES and ePWK are excluded from this comparison; their results are provided in supplementary material for completeness. All methods were calibrated to operate under comparable computational budgets, with method-specific numbers of posterior draws adjusted accordingly.

In dimensions $d=5$ and $d=10$, the tTHAMES method encounters difficulties when computing the intersection volume using simple Monte Carlo approximation: the uniform samples drawn within the ellipsoid often fail to fall inside the HPD-truncated intersection region, resulting in a computed volume of zero. To address this, the method employs a multi-step strategy, shifting the ellipsoid center from the posterior mean to the posterior mode and shrinking the radius of the initial ellipsoid. While this adaptive approach allows the algorithm to proceed, it highlights the challenging geometry of the Rosenbrock distribution in higher dimensions and the limitations of relying on a single global ellipsoid to capture such complex posterior structures.

Across all configurations, ECMLE consistently provides the most accurate and stable estimates, with values tightly concentrated around the true marginal likelihood ($Z=1$).  This demonstrates the effectiveness of ECMLE's adaptive ellipsoid covering in capturing the curved, banana-shaped geometry of the Rosenbrock distribution.

\section{Discussion}\label{sec:discus}
Similar to the importance sampling principle, the harmonic mean estimator approach offers a wide range of possibilities with equally widely ranging efficiencies. It is somewhat unfortunate that the first version defined in \cite{newton1994approximate} became the default version, despite immediate warnings from \citet{neal1994contribution} that it could produce highly unstable approximations in a large variety of cases. The contemporary generalization proposed by \citet{gelfand1994bayesian} did not have the same impact on the community. The realization by \citet{robert2009computational} that uniform distributions on high-density sets could be used, led to a renewed interest in the approach and the current paper provides a manageable approximation of HPD regions by a collection of non-overlapping ellipsoids that serves as a well-defined support. Appealing features of the approach include recycling MCMC simulations and not requiring the complete identification of all modal regions. Here, we studied the impacts of both the coverage level and the shape of the ellipsoids, but further calibrations should also be examined, first and foremost the impact of the parameterization of the parameter space used for the estimator \eqref{zhat}. Seeking this optimal parameterization is akin to finding a normalizing flow \citep{papamakarios2021} that minimizes the variance of the evidence estimator, provided derivations are light enough on the computing side. Further convergence assessment techniques could also be introduced, as for instance in using ECMLE based on different sets as control variates. Future extensions could also aim to enhance computational scalability and reduce variance in higher-dimensional settings, for example through dimensionality-reduction techniques or hybrid variance-minimization strategies.
 Finally, a formal study of the impact of the dimension $d$ of the parameter space on the choice of the level $\alpha$ could return a more principled choice of this level.


\begin{funding}
Dana Naderi acknowledges support from a scholarship awarded by the Ministère de l’Europe et des Affaires étrangères (MEAE).
Christian P. Robert is also affiliated with the University of Warwick, UK, and ENSAE-CREST, Palaiseau, and he is partly funded by the European Union under the GA 101071601, through the 2023-2029 ERC Synergy grant OCEAN. This work has also benefited from a support by Agence Nationale de la Recherche through the France 2030 with reference ANR-23-IACL-0008, on a PR[AI]RIE-PSAI chair.
\end{funding}


\begin{supplement}
\subsection*{Exact Evidence Computation of Rosenbrock Example} \label{RosSup}

Let 
\(
\bar{Y} = (\bar{Y}_1, \bar{Y}_2, \ldots, \bar{Y}_d)^{\top} \in \mathbb{R}^d
\)
denotes the sample mean of \( n \) observations.

\noindent
\textbf{Likelihood:}
\[
\bar{Y}_j \mid \theta \sim \mathcal{N}(\mu_j(\theta), \sigma^2 / n), \quad j = 1, 2, \ldots, d
\]
where
\[
\mu_1(\theta) = \theta_1, \quad \text{and} \quad \mu_j(\theta) = \theta_j + b_{j-1}(  \theta_{j-1}^2 - a_{j-1}), \quad j = 2, \ldots, d.
\]

\noindent
The full likelihood density is:
\[
p(\bar{Y} \mid \pmb{\theta}) = C \cdot \exp\left( -\frac{n}{2\sigma^2} \sum_{j=1}^{d} (\bar{Y}_j - \mu_j(\pmb{\theta}))^2 \right),
\]
where $C = \left( \frac{n}{2\pi\sigma^2} \right)^{d/2}$.

\noindent
\textbf{Prior:} Flat (improper) prior on all parameters: $\pi(\pmb{\theta}) \propto 1$.

\noindent
\textbf{Marginal likelihood:}
We compute $Z = C \cdot I$, where
\[
I = \int_{\mathbb{R}^d} \exp\left( -\frac{n}{2\sigma^2} \sum_{j=1}^{d} (\bar{Y}_j - \mu_j(\pmb{\theta}))^2 \right) d\pmb{\theta}.
\]

\noindent
\textbf{Change of variables:}
Define $\pmb{\phi} = g(\pmb{\theta})$ by $\phi_j = \mu_j(\pmb{\theta})$ for all $j$:
\[
\phi_1 = \theta_1, \quad \phi_j = \theta_j + b_{j-1}(\theta_{j-1}^2 - a_{j-1}) \quad \text{for } j = 2, \ldots, d.
\]
The Jacobian matrix $J = \frac{\partial \pmb{\phi}}{\partial \pmb{\theta}}$ is lower bidiagonal:
\[
J = \begin{pmatrix}
1 & 0 & 0 & \cdots & 0 \\
2b_1\theta_1 & 1 & 0 & \cdots & 0 \\
0 & 2b_2\theta_2 & 1 & \cdots & 0 \\
\vdots & \ddots & \ddots & \ddots & \vdots \\
0 & \cdots & 0 & 2b_{d-1}\theta_{d-1} & 1
\end{pmatrix}.
\]
Since $J$ is lower triangular with all diagonal entries equal to 1, we have $\det(J) = 1$ for all $\pmb{\theta} \in \mathbb{R}^d$.

\noindent
\textbf{Evaluation:}
With $|\det(J)| = 1$, the change of variables gives $d\pmb{\theta} = d\pmb{\phi}$, and the integral becomes:
\[
I = \int_{\mathbb{R}^d} \exp\left( -\frac{n}{2\sigma^2} \sum_{j=1}^{d} (\bar{Y}_j - \phi_j)^2 \right) d\pmb{\phi} = \prod_{j=1}^d \int_{-\infty}^{\infty} \exp\left( -\frac{n}{2\sigma^2}(\bar{Y}_j - \phi_j)^2 \right) d\phi_j.
\]
Each integral equals $\sqrt{2\pi\sigma^2/n}$, so $I = \left( \frac{2\pi\sigma^2}{n} \right)^{d/2}$. Therefore:
\[
Z = C \cdot I = \left( \frac{n}{2\pi\sigma^2} \right)^{d/2} \times \left( \frac{2\pi\sigma^2}{n} \right)^{d/2} = 1.
\]
\newpage
\subsection*{Additional Results of Rosenbrock Example}
\label{appendix_b}
\begin{figure}[H]
\includegraphics[width=\textwidth]{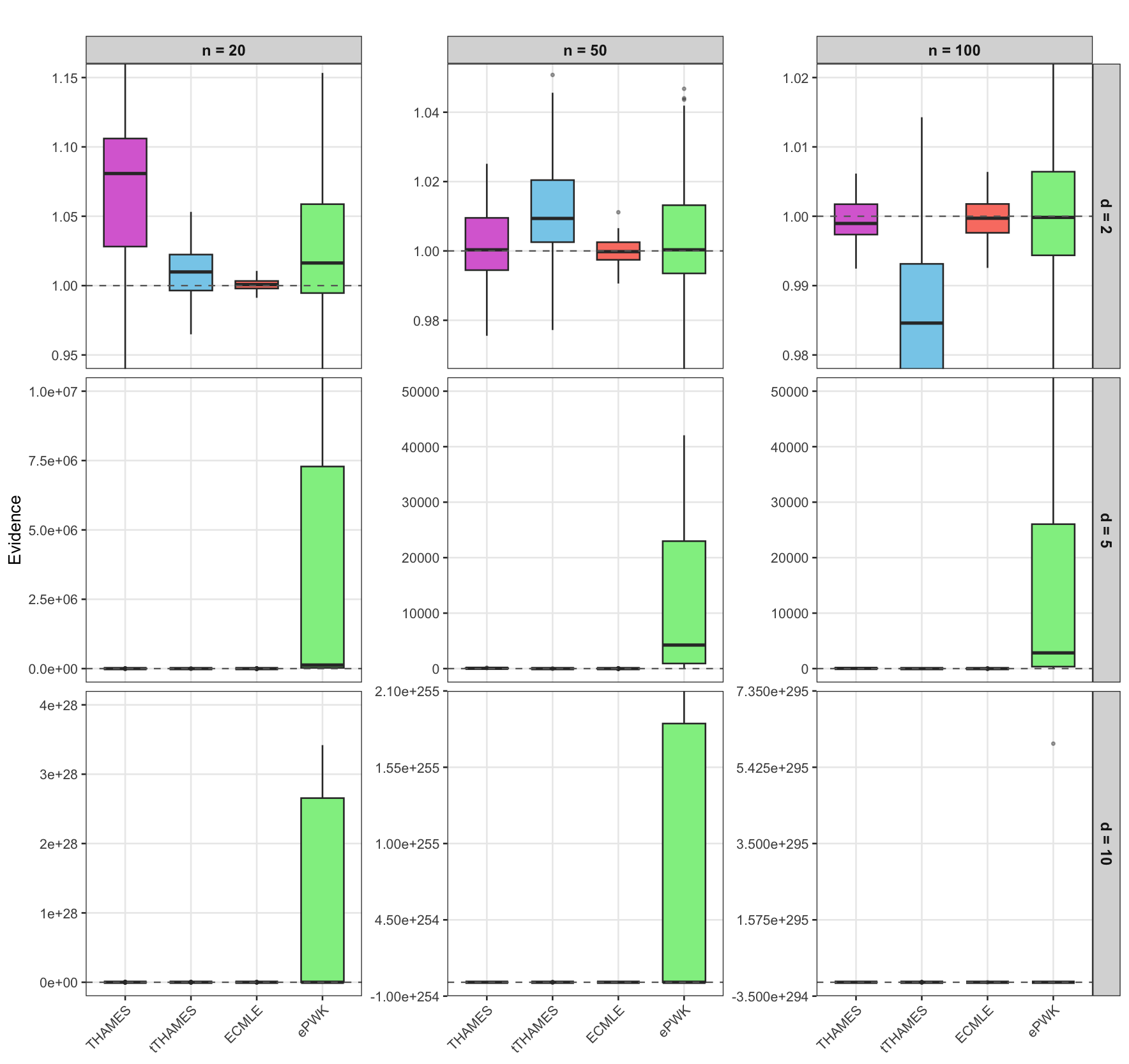}
        \caption{({\bf Example 3}) Boxplots of marginal likelihood estimates for the Rosenbrock distribution posterior in dimensions $d \in \{2,5,10\}$ and sample sizes $n \in \{20,50,100\}$, based on $M=100$ independently generated datasets. The y-axis scale is adjusted to show that ePWK produces estimates far outside the acceptable range, particularly in higher dimensions. For each dataset, all methods use the same posterior sample, but with method-specific numbers of draws chosen so that all methods operate under a comparable computational budget.
 }
    \label{fig:Rosen_ePWK}
\end{figure}
\begin{figure}[h!]
\includegraphics[width=\textwidth]{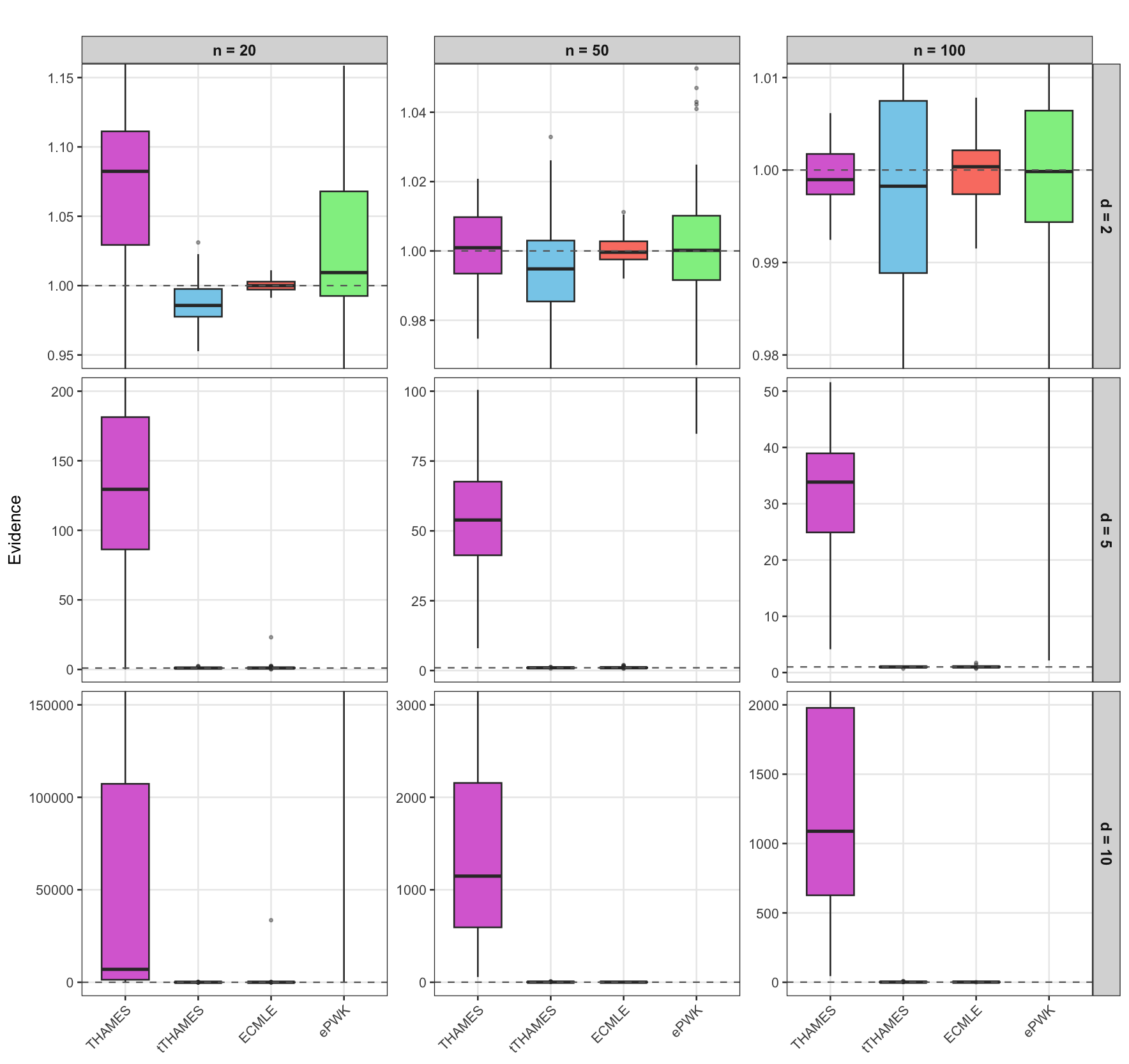}
        \caption{({\bf Example 3}) Boxplots of marginal likelihood estimates for the Rosenbrock distribution posterior in dimensions $d \in \{2,5,10\}$ and sample sizes $n \in \{20,50,100\}$, based on $M=100$ independently generated datasets. The y-axis scale is adjusted to show that THAMES produces estimates far outside the acceptable range, particularly in higher dimensions. For each dataset, all methods use the same posterior sample, but with method-specific numbers of draws chosen so that all methods operate under a comparable computational budget.
 }
    \label{fig:Rosen_THAMES}
\end{figure}

\end{supplement}
\end{document}